\documentclass[reprint, superscriptaddress, nofootinbib, floatfix]{revtex4-1}

\usepackage[utf8]{inputenc}
\usepackage{graphicx,xcolor,amsmath,amssymb,bm,caption,subcaption}
\usepackage{hyperref}
\hypersetup{
    colorlinks=true,
    linkcolor=black,
    filecolor=black,      
    urlcolor=blue,
    citecolor=black
    }

\begin{document}

\title{Applications of Persistent Homology in Nuclear Collisions}
\author{Greg Hamilton}
\affiliation{Institute for Condensed Matter Theory, Department of Physics, University of Illinois at Urbana-Champaign, Urbana, IL 61801, USA}
\author{Travis Dore}
\affiliation{Illinois Center for Advanced Studies of the Universe, Department of Physics, University of Illinois at Urbana-Champaign, Urbana, IL 61801, USA}
\author{Christopher Plumberg}
\affiliation{Illinois Center for Advanced Studies of the Universe, Department of Physics, University of Illinois at Urbana-Champaign, Urbana, IL 61801, USA}
\affiliation{Natural Science Division, Pepperdine University, Malibu, CA 90263, USA}
\date{\today}

\begin{abstract}
    We introduce a novel set of observables associated to the rapidly developing field of \textit{persistent homology} for the quantitative characterization of nuclear collisions and their evolution. Persistent homology allows for the identification of topological and homological characteristics of distributions in multi-dimensional spaces. We demonstrate here how to apply the toolset of persistent homology to the extraction of novel clustering signatures and the identification of long-range flow correlations in the particle production process of nuclear collisions.
\end{abstract}

\maketitle

\section{Introduction}

The field of relativistic nuclear collisions exists to explore the properties of quantum chromodynamic (QCD) matter at asymptotically large temperatures and densities \cite{Jacobs:2004qv, Baym:2016wox}.  In addition to creating a novel phase of deconfined matter known as the \textit{quark-gluon plasma} (QGP) \cite{Shuryak:2014zxa}, nuclear collisions provide insights into the equation of state of nuclear matter \cite{Braun-Munzinger:2008szb, Ratti:2018ksb, Noronha-Hostler:2019ayj}, conjectured topological characteristics of the QCD vacuum \cite{Kharzeev:2004ey, Kharzeev:2007jp, Kharzeev:2020jxw}, input into models of neutron star structure and mergers \cite{Tan:2020ics, Dexheimer:2020zzs, Most:2021zvc}, and much more \cite{PHENIX:2018lia, HADES:2019auv, Bluhm:2020mpc}. To date, a vast number of observables have been used to probe nuclear collisions, including particle multiplicities \cite{ALICE:2015juo, ALICE:2019hno}, $p_T$ and rapidity distributions \cite{CMS:2011aqh, STAR:2013gus}, anisotropic flow \cite{ATLAS:2013xzf, ALICE:2014dwt, STAR:2015mki, ALICE:2017kwu, ALICE:2018yph}, jet-quenching \cite{STAR:2009ngv, ALICE:2013dpt, CMS:2013qak}, fluctuations and correlations of conserved charges \cite{ALICE:2012xnj, STAR:2017tfy}, interferometry and femtoscopy \cite{ALICE:2011dyt, PHENIX:2015jaj}, and a litany of others \cite{NA49:2002pzu, CMS:2012bms, ALICE:2018gif, ALICE:2021hjb}.

\begin{figure}
    \centering
    \includegraphics[width=.65\linewidth]{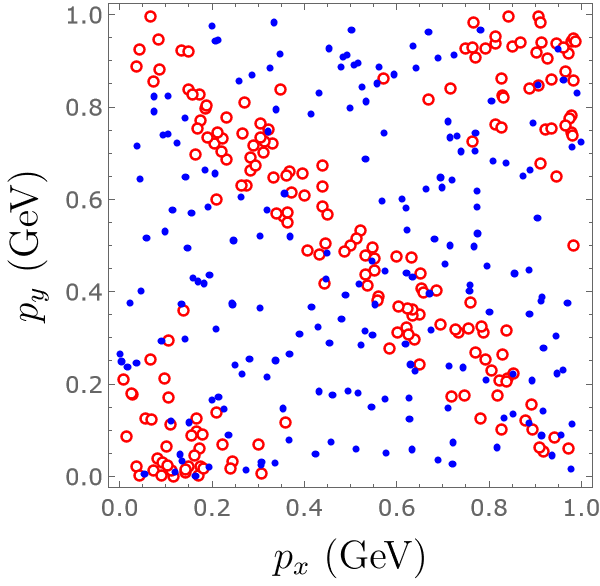}
    \caption{A toy example of a point cloud in two-dimensional momentum space. Each point represents a particle emitted from an event, where the point cloud distributions of the solid blue points clearly differ from that of the open red circles.  These are the sort of qualitative features which persistent homology is designed to access. }
    \label{F1}
\end{figure}

What all of these observables have in common is that they are constructed from \textit{point clouds}. A point cloud, as defined in this paper, is simply a distribution of points in some $d$-dimensional space (cf. Fig.~\ref{F1}). Point clouds generically arise as finite samples from an underlying continuous distribution and may reflect non-trivial topological structure present in the latter. In the case of nuclear collisions, each collision (or ``event") emits a number of particles which are detected, and whose three-momenta can be measured experimentally. The fundamental insight of this paper is to treat these emitted particles as a point cloud in momentum space, where each particle exists as a point with three-dimensional coordinates given by its three-momentum $\vec{p}$ as measured by the detector. One therefore has access experimentally to an \textit{ensemble} of point clouds which can be mined for insights into the underlying dynamics and properties of the nuclear collisions which produced them.

\begin{figure*}
    \centering
    \includegraphics{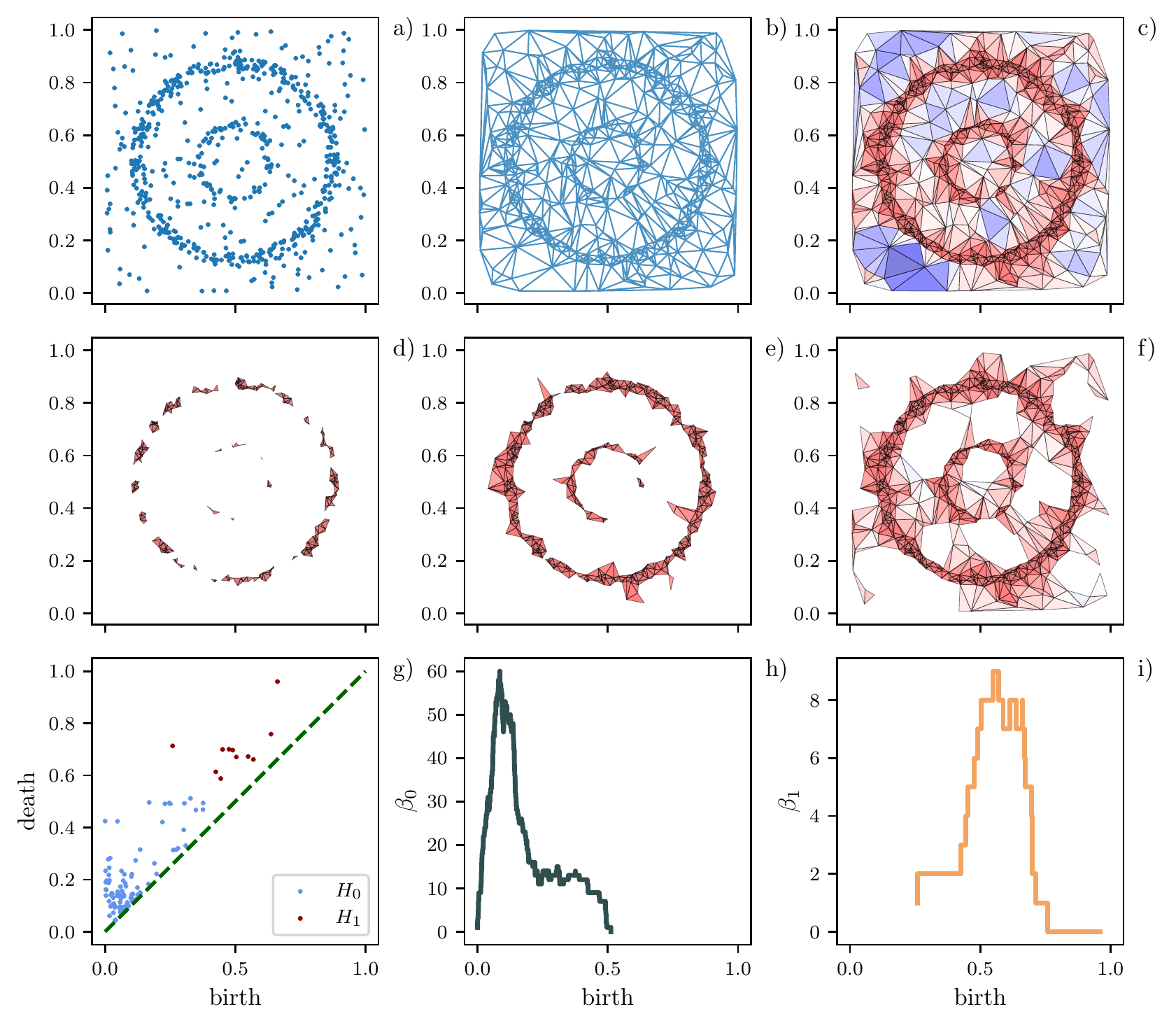}
    \caption{PH pipeline used in this work. a) Random point cloud with multiplicity $n\approx 2000$. b) The Delaunay triangulation, note the prevalence of long ``sliver" triangles at the boundary. c) The DTFE with the color of any triangle roughly scaling with average density of its vertices: warm (red) colors denote high density regions, while cool (blue) colors denote low density regions. The precise determination of the densities is described in the main text. d-f) Superlevel set of the density field whereby $(95\%,75\%,50\%)$ of the points have density lower than the threshold $\varepsilon$. g) The resultant persistence diagram for $H_{0}, \, H_{1}$. h) Betti curve for $0$D. i) Betti curve for $1$D.}
    \label{fig:ph_pipeline}
\end{figure*}

The toolset of persistent homology (PH) has been designed for exactly this purpose. PH has developed rapidly in recent years as one of the foremost techniques for non-parameterically identifying important topological characteristics of large datasets, including point cloud distributions. PH is best suited to identifying topological or homological features of a given point cloud, including clustering, bubbles, filaments, holes, walls, and so on. It has been applied in a vast number of other disciplines, including the description and evolution of cosmic structure \cite{Xu:2018xnz, Biagetti:2020skr, Wilding:2020oza}, Bose-Einstein condensates \cite{10.21468/SciPostPhys.11.3.060}, phase diagrams \cite{PhysRevE.93.052138} and even the assembly and disassembly of multispecies ecological systems \cite{angulo2021coexistence}. 
While PH yields access to topological features at varying degrees of scales, it also implicitly probes multi-order correlational structure. Indeed, PH has strong connections to robust results in Morse theory and quantifies large scale structure much like the Minkowski functionals for convex bodies, which can be interpreted in terms of integrated connected correlation functions at all orders \cite{Schmalzing_Buchert_1997,Schmalzing_1997,wiegand2014direct}. Further, deep connections to the Gauss-Bonnet theorem and the Euler characteristic \cite{Schmalzing_Buchert_1997} render PH a promising extension of traditional statistical methods for analyzing discrete point clouds, and make it a natural candidate for developing new ways of probing nuclear collisions. \par 
Our topological approach complements and extends a recent explosion of machine learning techniques in high-energy physics, most prominently tools utilizing neural networks like graph neural networks \cite{thais2022graph}. What is more, persistent homology is easily folded into a machine learning pipeline and can help identify the topological structure of information as it passes through the layers of a neural network \cite{adams2021topology}.  Developing a formalism for applying PH to nuclear collision phenomenology therefore broadens the possible avenues for connecting it to the field of machine learning.

\par 
In this paper, we show how the PH toolset can be applied to the description of nuclear collisions and the momentum-space point clouds they produce. For illustrative purposes, and as a proof-of-concept, we use it to introduce novel ways of quantifying both clustering effects and anisotropic collective flow in realistic nuclear collision simulations. Our goal is to exhibit some of the ways in which PH could be leveraged to provide new insights into the evolution and phenomenology of nuclear collisions.

The outline of this paper is as follows. In Sec. \ref{sec:overview_ph}, we provide a brief review of the main results, tools, concepts, and techniques employed in PH, and summarize the pipeline by which we analyze a given point cloud in this work. Then, in Sec. \ref{sec:observables}, we discuss specifically how we apply these results in the context of nuclear collisions, and introduce several PH observables which are designed to probe familiar nuclear collision phenomenology. Secs. \ref{sec:demo} and \ref{sec:results} present an illustrative proof-of-concept for the application of our novel methods to realistic nuclear collisions, based on established simulation packages for modeling the real-time evolution of these systems. Finally, we conclude with an assessment of the prospects for applying PH in nuclear collisions and suggest further questions which our novel approach could help to clarify.

\section{Concepts of Persistent Homology}\label{sec:overview_ph}

In this section we present an overview of PH, focusing especially on its use as a tool to investigate higher-order correlational structures.  We begin with a brief description of PH in general and introduce the pipeline we use to apply PH to simulated nuclear collision data.  For the sake of clarity, we illustrate the most important concepts from our pipeline using a toy dataset which contains artificial topological structures that are naturally probed by PH.  

\subsection{PH overview}
At a high-level, PH quantifies how topology persists with respect to a variational parameter. This variational parameter introduces a nesting (or \textit{filtration}) of topological spaces.  Each of these topological spaces is typically triangulated, yielding what is formally known as a \textit{simplicial complex} (an example is shown in Fig.~\ref{fig:ph_pipeline}b).  In turn, by making use of the techniques of simplicial homology \cite{ghrist2008barcodes}, each simplicial complex in a filtration can be connected with a sequence of homology groups. The key insight of persistent homology is to track how these homology groups change as a function of the filtration parameter, thus providing insight into the topological structure reflected in the point cloud. The output of this process can be usually represented by a plot known as a \textit{persistence diagram}, and may be analyzed further by means of various observables designed to isolate relevant features of a given point cloud ensemble.

There are several excellent reviews \cite{otter2017roadmap,carlsson2020persistent} which we encourage the reader to consult for further discussion of PH in general.  In the remainder of this section we provide a description of the specific PH pipeline which we apply to the output of simulated nuclear collisions.  The pipeline consists of three main steps, described below, with technical details of each step deferred to Appendix \ref{sec:app_dtfe}.

\subsection{Summary of PH pipeline}

We illustrate the results of our pipeline when applied to a toy dataset in Fig.~\ref{fig:ph_pipeline}.  The dataset itself is a 2D Euclidean point cloud, denoted $X$, and depicted in Fig.~\ref{fig:ph_pipeline}a. While $X$ is a point cloud (and therefore has trivial topological structure), it was sampled from a continuous distribution consisting of two concentric annuli, and clearly reflects the non-trivial topology of the latter.  Some additional points have also been added as noise.  The topologically non-trivial loop structures exhibited by this toy dataset can be characterized by following three main steps: (1), performing a Delaunay triangulation and associated field estimation; (2), conducting a superlevel set filtration; and (3), identifying homological features of interest and evaluating relevant observables which quantify these features.  We now discuss each of these three steps in greater detail.

\subsubsection{Delaunay Triangulation and Field Estimation}

First, we generate a Delaunay triangulation of $X$, as shown in Fig. \ref{fig:ph_pipeline}b. The Delaunay triangulation is a non-parametric triangulation such that, in Euclidean space, no points appear in the circumcircle interior of any triangle \cite{fortune1995voronoi}.   The boundary of the Delaunay triangulation is called the \textit{convex hull} of the point cloud.

After constructing the triangulation, we use a technique known as Delaunay Triangulation Field Estimation (DTFE) to define a density field $f(x)$ on the points $x\in X$ \cite{van2009geometry}. The DTFE assigns to each point a density which depends upon the area of adjacent triangles and thus correlates closely with the density of neighboring points in the vicinity (cf. Appendix \ref{sec:app_dtfe} for more details). The results of the DTFE are depicted in Fig. \ref{fig:ph_pipeline}c. The density field has been normalized to the range $[0,1]$, although we use the unnormalized density when analyzing nuclear collisions below. The colors of the triangles correspond roughly to the average $f(x)$ of the vertices of the triangle. The two annuli reflect regions of higher density, and are thus denoted by redder colors; lower density regions are colored blue.
    
\subsubsection{Superlevel set filtration}

After constructing the Delaunay triangulation and performing the field estimation, we introduce a variational parameter $\varepsilon$ and consider the set of points $L^{+}_{\varepsilon}:= \{x| f(x) \ge \varepsilon\}$, which is the collection of points in the cloud where the density is greater than or equal to $\varepsilon$. This kind of set is referred to as a \textit{superlevel set} of the density field $f(x)$.  By definition, we have $L^{+}_{\varepsilon} \subseteq L^{+}_{\varepsilon'}$ whenever $\varepsilon' \le \varepsilon$.  $L^{+}_{\varepsilon}$ is thus a filtration, where $\varepsilon$ plays the role of the filtration parameter.

For a given value of $\varepsilon$, a simplicial complex $K_{\varepsilon}$ can be constructed in the following way.  First, the vertices (or \textit{0-simplices}) in $K_{\varepsilon}$ are identified with the points $x \in L^{+}_{\varepsilon}$.  Once the 0-simplices have been specified, the $1$-simplices are identified as the edges in the triangulation that connect two vertices.  Similarly, any triangles are identified as 2-simplices, tetrahedra as 3-simplices, and so on. Thus, for example, any triangle which is formed by three edges (or 1-simplices) in $K_{\varepsilon}$ is ``filled in" and included in $K_{\varepsilon}$ as a $2$-simplex.  The same applies to higher-order simplices.

Figs. \ref{fig:ph_pipeline}d-f show $K_{\varepsilon}$ for three values $\varepsilon_{d} \ge \varepsilon_{e} \ge \varepsilon_{f}$.  The thresholds $\varepsilon_{d/e/f}$ in the Figure correspond to points in, respectively, the top 5\%, the top 25\%, and the top 50\% of the densities provided by the DTFE.  Note that each simplicial complex is a subcomplex of the subsequent complex. Note also that the inner annulus is a fully connected component in Fig.~\ref{fig:ph_pipeline}d, and that the ``loop" structure is clearly formed.  A particular filtration thus generates a corresponding sequence of nested simplicial complexes which can reflect topological structures underlying the original point cloud.

\subsubsection{Identify homological features}


Once a filtration and sequence of simplicial complexes have been defined, the associated homology groups follow immediately. For any $\varepsilon$, we compute the simplicial homology of $K_{\varepsilon}$ to obtain a direct sum of the homology groups (vector spaces) reflected in $K_{\varepsilon}$.  The rank of each homology group is known as its \textit{Betti number}, $\beta_{i}$, where $i$ denotes the dimension. Intuitively, $\beta_{i}$ counts the number of homological features of dimension $i$: $\beta_{0}$ is the number of connected components, $\beta_{1}$ is the number of non-homologous loops, and so on. In Fig. \ref{fig:ph_pipeline}h-i we show the Betti numbers $\beta_{0}, \, \beta_{1}$, respectively, as a function of $\varepsilon$; these plots are known as \textit{Betti curves}. The nesting of simplicial complexes ensures a common basis to track which homology groups persist through the filtration $\{\varepsilon\}_{\varepsilon\in [0,1]}$.

Finally, by tracking the homology groups through the filtration one obtains a \textit{persistence diagram} (PD), shown in Fig. \ref{fig:ph_pipeline}g. The abscissa (``birth" or $b$-axis) is the filtration value at which a homological feature first appears, while the ordinate (``death" or $d$-axis) indicates at which filtration value the same homological feature vanishes. The blue markers, denoted $H_{0}$ to indicate the $0$D homology group, correspond to the point cloud's connected components. $H_{1}$ denotes the 1D homology group and represents the loops (formally, \textit{cycles}) present in the point cloud. The distribution of points in the PD thus serves as a ``fingerprint" for the topology of the point cloud: the farther points in the PD are from the diagonal, the ``longer-lived" the corresponding homological feature. We define the difference $d-b$ as the ``lifetime" of the homological feature. Thus, long-lived features correspond to large-scale structure, while short-lived features correspond to noise and local curvature \cite{bubenik2020persistent,adams2021topology}.

The example given here is identical to the PH pipeline we apply to our simulated collision data, save a few caveats. In this work we consider point clouds in $(\phi,y)$ coordinates, where \begin{align}
    p_{x} &= p_{T}\cos \phi, \\
    p_{y} &= p_{T}\sin \phi, \\
    p_{z} &= m_{T}\sinh y,
\end{align} and $y$ is the rapidity $m_T \equiv \sqrt{m^2 + p_T^2}$. For consistency, we must have periodic boundary conditions in the $\phi$ direction, which complicates the Delaunay triangulation and tends to generate spurious edge effects induced by the subsequent DTFE. To avoid these complications, we impose a rapidity cut $|y|\le 2$, which implies that our observables outlined below are defined within this rapidity interval. We discuss this further in Appendix \ref{sec:app_dtfe}.

Furthermore, the pipeline we present here is not the only way PH could be applied to nuclear collisions.  For instance, although we have employed $(\phi,y)$ coordinates in this work, one could also consider analyzing particle spectra in three dimensions using coordinates $\vec{p} = (p_x,p_y,p_z)$. Similarly, we use a density-based filtration previously applied in the context of cosmological models for galactic morphology \cite{Pranav_Edelsbrunner_van}, but there are several alternative ways to perform PH on a point cloud as well, as we discuss in Appendix \ref{sec:app_diff_PH}.  We leave a more in-depth analysis of these various possibilities and extensions to future work.

\section{Observables}\label{sec:observables}

While in principle the PD contains a great deal of information about a single point cloud's persistent topology, in practice it can be difficult to analyze its \textit{statistical} properties for an ensemble of point clouds.  For this reason, it is often desirable to introduce a scalar quantity or functional (known as a \textit{topological summary}), derived from the PD, for which one can readily formulate and quantify relevant statistical properties.  Many ways to do this have been discussed in the literature, including persistence landscapes \cite{bubenik2020persistence}, persistence images \cite{adams2017persistence}, and statistics on the birth, death, and lifetime distributions \cite{atienza2020stability}. Each topological summary yields unique insights into fluctuations of the intrinsic topology of an ensemble of point clouds. In this work we focus on four such summaries, two of which actually incorporate more information than the PD alone.  These summaries thus provide observables which can be applied to an ensemble of nuclear collisions.

\subsection{Fractal Dimension}
The first observable we consider is known as the \textit{fractal dimension}.  While the birth and death distributions of homological features are interesting in their own right, the \textit{lifetime} distribution in particular quantifies how persistent topological features are in the underlying point cloud. Moreover, we can study how the lifetime distribution changes as the size (or \textit{multiplicity}) of a point cloud representing some dynamical process increases. The scaling of persistent topology with respect to multiplicity thus gives rise to a notion of fractality which we can quantify, and which was formally described in Ref. \cite{Jaquette_Schweinhart_2020}.\par 
Given a point cloud with multiplicity $n$ and its corresponding PD, let PD$_{i}$ denote the restriction to homological features of dimension $i$. Then let $E^{i}_{\alpha}:= \sum_{(b,d)\in PD_{i}}(d-b)^\alpha$ denote a sum of powers of the lifetimes in PD$_{i}$. The fractal dimension is then defined as \begin{align}\label{eq:beta}\dim PH_{i}: = \frac{\alpha}{1-\beta},\, \beta = \lim_{n\to \infty} \frac{\log \langle E^{i}_{\alpha}\rangle  }{\log n}.\end{align} Here $\langle \cdot \rangle$ denotes averaging over persistence diagrams with the same multiplicity $n$ \cite{Jaquette_Schweinhart_2020}. Intuitively, Eq. \eqref{eq:beta} measures how the sum of powers of the lifetimes scales with multiplicity. Small values of $\alpha$ emphasize the small lifetime features (e.g., local clustering), while large $\alpha$ probes more global features. The homological fractal dimension defined here has close connections to the box-counting and correlation dimensions, and has been explored in the context of identifying critical exponents and fractal dimensions for dynamical processes \cite{Jaquette_Schweinhart_2020}. Moreover, a notion of fractal dimension has recently been explored in the context of jet classification \cite{Davighi_Harris_2018}.

\subsection{Betti Curves}
As noted above, the Betti number $\beta_{n}$ is simply the rank of the $n$th homology group and reflects the number of important topological features of dimension $n$ at a given stage in the filtration. While the Betti curve is insensitive to the relative lifetimes of homological features, it does give an indication of the point in the filtration at which homological features are likely to exist. The structure of the Betti curve, in particular, the maximum, has recently been used in the context of identifying phase transitions in quantum many-body systems \cite{olsthoorn2021persistent}. As we explore in Sec. \ref{sec:results}, the Betti curve also serves as a cluster distribution function with respect to scale; i.e., yielding the (unnormalized) probability of $m$ clusters at scale $\varepsilon$. These cluster distribution functions form an important set of observables in cosmological studies of galactic morphology \cite{hamaus2014modeling}, leading us here to consider their relevance for nuclear collisions as well.

\begin{figure}
    \centering
    \includegraphics[width=.8\linewidth,trim={2cm 2cm 2cm 4cm},clip]{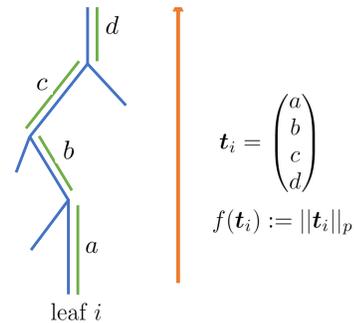}
    \caption{A pictorial representation of a merge tree, or dendrogram. The filtration from leaves to root runs up the page, with the height of a leaf corresponding to the value at which a hadron appears in the filtration. The length of the green bars from the lowest leaf to the root indicate the lifetimes of the clusters that include the lowest leaf. The set of these lifetimes for each leaf $i$ form the vector $\bm{v}_{i}$.}
    \label{fig:merge_tree}
\end{figure}

\subsection{Cluster Entropy} While the 0D Betti curve is a useful indicator for the distribution of clusters with respect to filtration value, the Betti number is insensitive to the multiplicity of individual clusters. For instance, given a point cloud of $n$ points with three clusters, the Betti number $\beta_{0}$ does not distinguish between three clusters with equal numbers of $n/3$ points each, versus one cluster with $n-2$ points and the other two clusters with one point each.\par 
To access this cluster multiplicity information, we exploit the fact that our density-based filtration amounts to a parameterized hierarchical clustering scheme due to the Delaunay triangulation. This hierarchical clustering is quite similar to the clustering scheme used to identify jets \cite{greenberg2021exact}; indeed, some collinear, infrared-safe jet clustering algorithms also make use of the Delaunay triangulation in $(\phi,y)$ space \cite{cacciari2006dispelling}. The output of hierarchical clustering is an object known as a \textit{dendrogram} (or \textit{merge tree}), wherein the lengths of branches between merges indicate the filtration interval in which a given cluster exists. The number of leaves of a branch yields the multiplicity of the cluster at that filtration level. Fig. \ref{fig:merge_tree} shows a small illustrative example; note that the height of the leaves are non-uniform because, in our filtration, the leaves appear at values related to the density.  The dendrogram thus provides a way of quantifying the distribution of cluster multiplicities as a function of filtration.

Given the number of points in each cluster as a function of the filtration parameter, we introduce a novel observable which we refer to as the \textit{cluster entropy}, which acts as a topological summary of agglomerative clustering. Given a point cloud of multiplicity $n$, at each filtration level $\varepsilon$ we define
\begin{align} H(\varepsilon) = -\sum_{C_{i} \in \mathcal{C}(\varepsilon)}p_{i}\log p_{i},
\end{align}
where $i \in \mathcal{C}(\varepsilon)$ is the set of clusters at $\varepsilon$ and $p_{i} = |C_{i}|/n$. The cluster entropy defined here is in effect the Shannon entropy \cite{amigo2018brief} of the clusters when treated as a probability distribution, and indicates the degree of ``mixedness" in the distribution of points amongst clusters. This statistic naturally generalizes to the Renyi and Tsallis entropies, which explore the ``rarity" or heavy tails of a distribution \cite{amigo2018brief}.

\subsection{Local Clustering Statistics}

A particularly important phenomenon frequently studied in nuclear collisions is that of \textit{local clustering}, which can arise in a variety of contexts, including Bose-Einstein correlations \cite{Zajc:1986sq}, the QCD critical point \cite{Shuryak:2018lgd, Shuryak:2019ikv, DeMartini:2020anq}, jet identification \cite{Cacciari:2008gp}, and $n$-body correlations arising from collective flow \cite{Heinz:2013th}.  By ``local clustering," we mean any significant deviation from a uniformly distributed point cloud which may be correlated with position within the point cloud.   In this sense, local clustering provides a generalized notion of ordinary clustering, which implies deviations from a uniform distribution but need not specify where the clustering takes place.  Since we wish to characterize non-uniform point cloud distributions using PH in a way which \textit{can} depend on the specific region of momentum space (e.g., for anisotropic flow), it is therefore essential to have a way of quantifying local clustering as well.

However, while persistence diagrams provide structural statistics on point clouds, PH alone does not retain information regarding other, non-topological degrees of freedom, such as whether or not clustering behavior is more likely in one part of the point cloud than another. While PH has been used to identify anisotropy in point cloud distributions and to quantify local curvature \cite{bubenik2020persistent}, PH does not explicitly retain positional degrees of freedom. 

However, the form of PH we pursue in this work comes with an object that does retain positional degrees of freedom: the dendrogram. Each leaf in a dendrogram corresponds to a point in the point cloud, and the lengths of branches between merges indicate how long a cluster persists before being merged into another cluster. To each leaf (point) we identify a vector $\bm{t}_{i}$, the components of which are the lengths of the branches along the path from the leaf $i$ to the root of the dendrogram. We depict a simple dendrogram in Fig. \ref{fig:merge_tree}, wherein the green bars denote the relevant branches from the leaf $i$ (bottom of the plot) up to the root. Every leaf in the dendrogram therefore has a corresponding vector $\bm{t}_{i}$.  Taking the $p$-norm of each $\bm{t}_{i}$ then yields a novel statistic $f(\bm{t}_{i}) := ||\bm{t}_{i}||_{p}$ (which we refer to as the leaf's \textit{$p$-norm}) on the point cloud that reflects the local clustering statistics: for large $p$ the $p$-norm emphasizes large scale clustering, while small $p<1$ (technically a seminorm) targets local clustering and local curvature. Here the $p$-norm for $\bm{t}_{i} = (t_{0},\ldots, t_{m})$ is defined as \begin{align}
    ||\bm{t}_{i}||_{p} = \left( \sum_{j}t_{j}^{p} \right)^{1/p}.
\end{align} This novel clustering statistic has to our knowledge not appeared in the topological data analysis literature, though recent approaches to merge trees that incorporate higher-dimensional homological information (known as decorated merge trees) have touched on similar ideas \cite{Pegoraro_2022,pont2021wasserstein,curry2022decorated}. Given the importance of anisotropy in higher-order correlation functions in the context of flow, we see the local clustering statistics observable as a reasonable step towards bridging the gap between persistent homology and traditional correlational metrics in nuclear collisions.

We thus have introduced four observables -- the fractal dimension, the Betti curves, the cluster entropy, and the $p$-norm associated to dendrogram leaves -- which can be readily applied to the analysis of nuclear collisions. These observables are designed to characterize different aspects of point clouds typically produced in nuclear collisions. In the next section we discuss the simulation framework to which these observables will be applied.

\begin{figure}
    \centering
    \includegraphics{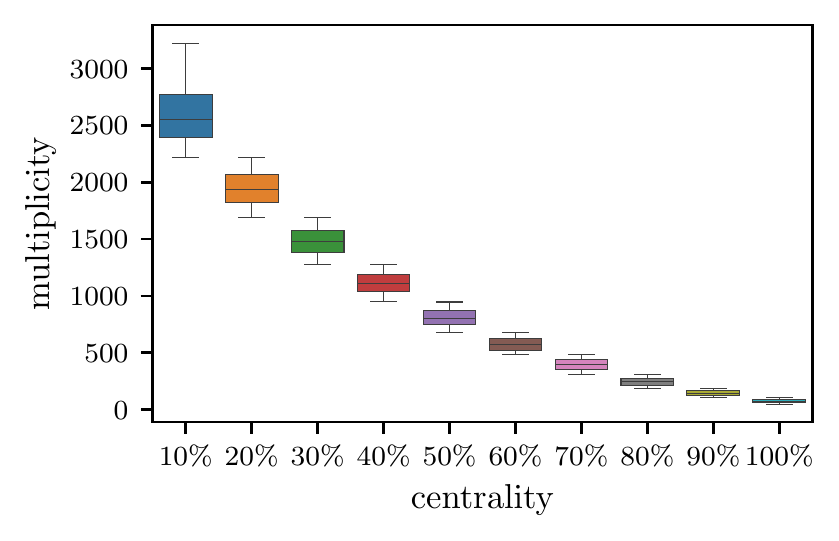}
    \caption{Box plot showing the distribution of multiplicities as a function of centrality class.}
    \label{fig:centrality}
\end{figure}

\section{Nuclear Collision Simulations}\label{sec:demo}
We now present our approach to generating realistic simulations of Pb+Pb collisions at LHC energies.  In addition to applying PH to the simulated Pb+Pb events themselves, it is also crucial to establish a suitable background as a reference against which to compare any proposed signals. Clearly for PH a suitable background should also account for the intrinsic topology of the ambient space in which a point cloud is situated, such as the cylindrical topology of the $(\phi,y)$ coordinate system.  As noted previously, the periodicity in the $\phi$ coordinate introduces some technical complications which we discuss more fully in Appendix \ref{sec:app_dtfe}. Below we discuss the details of our simulation framework and describe how we construct backgrounds for the PH observables we consider here.

\begin{figure*}
    \centering
    \includegraphics{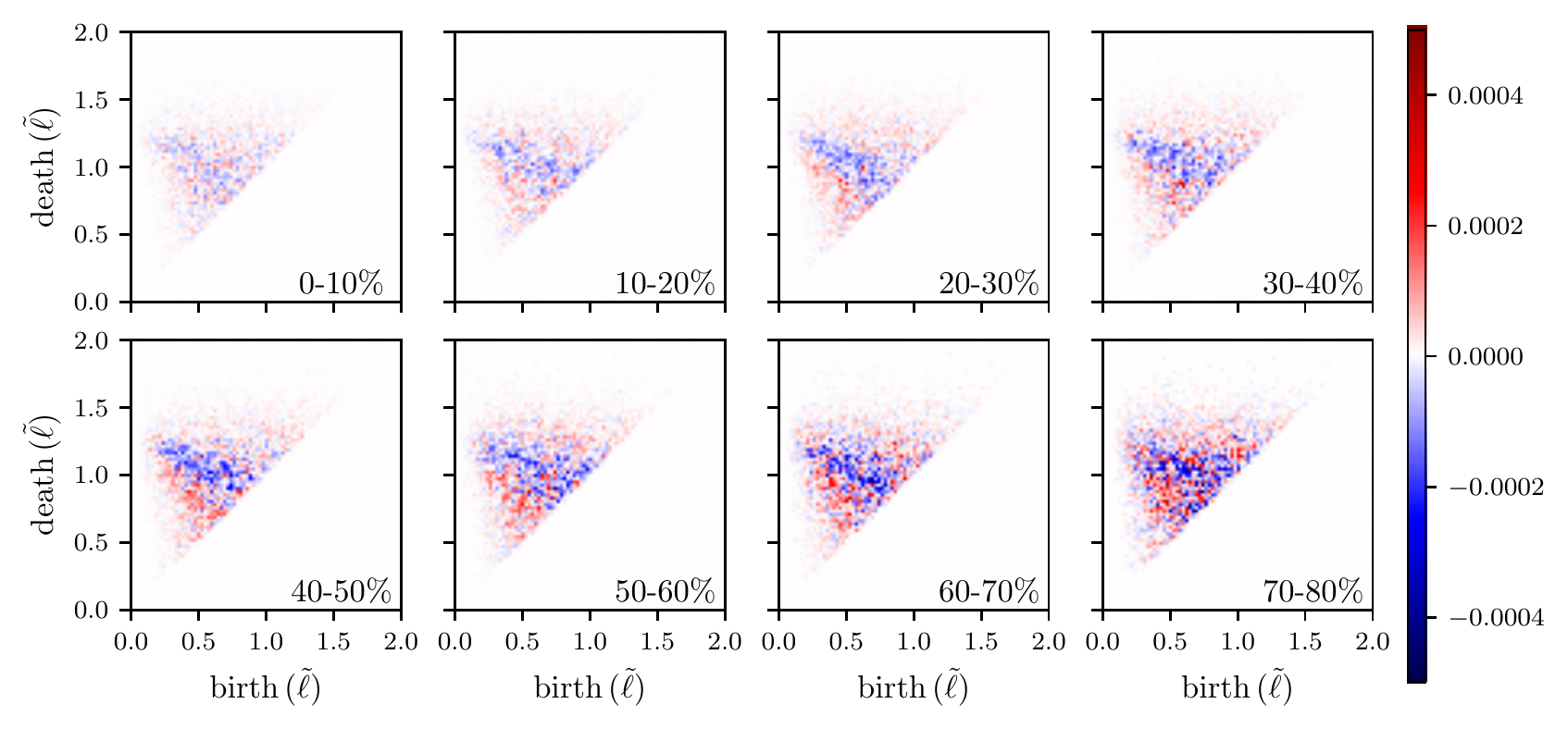}
    \caption{Differences in 0D PDs for each of the centrality classes. Note that the abscissa and ordinate are in terms of $\tilde{\ell}$.}
    \label{fig:pers_diag_0}
\end{figure*}

\begin{figure*}
    \centering
    \includegraphics[]{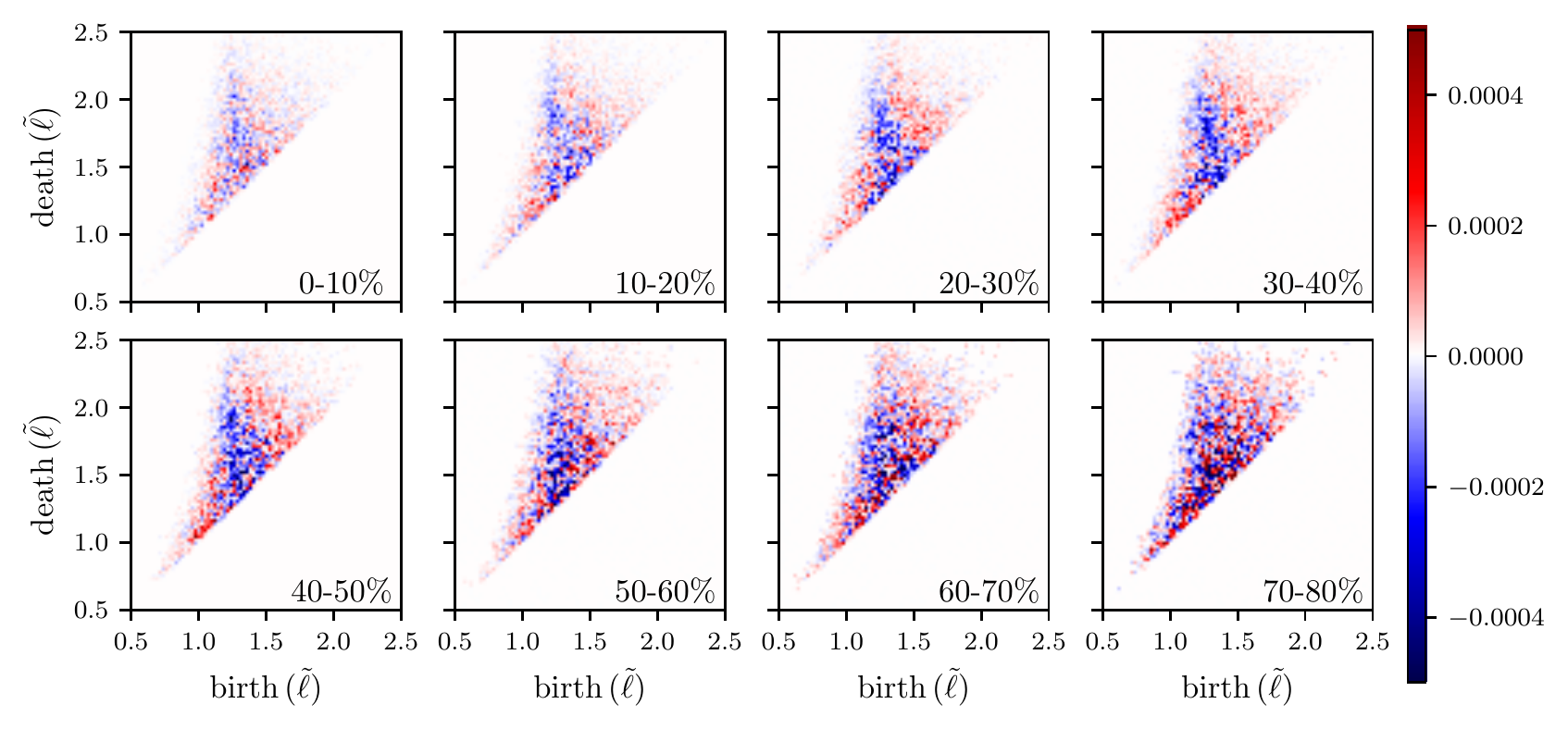}
    \caption{Differences in 1D PDs for each of the centrality classes. Note that the abscissa and ordinate are in terms of $\tilde{\ell}$.}
    \label{fig:pers_diag_1}
\end{figure*}

\subsection{Hydro simulations}

In this study, we have considered an ensemble of 10000 Pb+Pb collision events at $\sqrt{s_{NN}} = 2.76$ TeV, simulated using the Duke Bayesian tune of the iEBE-VISHNU framework to LHC p+Pb and Pb+Pb data \cite{Shen:2014vra, Moreland:2018gsh, Bernhard:2019bmu}.  This approach couples together T$_{\rm R}$ENTo initial conditions \cite{Moreland:2014oya} with a conformal, pre-hydrodynamic free-streaming phase \cite{Broniowski:2008qk, Liu:2015nwa}, a boost-invariant hydrodynamic phase \cite{Song:2007ux, Shen:2014vra} using the Denicol-Niemi-Molnar-Rischke (DNMR) formalism \cite{Denicol:2012cn}, and a hadronic afterburner UrQMD \cite{Bass:1998ca, Bleicher:1999xi}. We use the maximum-likelihood parameters  \cite{Bernhard:2019bmu} for the transport coefficients.  The hydrodynamic evolution is terminated at a hypersurface of constant $T_{FO} = 150$ MeV, corresponding to an energy density of $e_{FO} \approx 0.265 \text{ MeV}/\mathrm{fm}^3$.  After the hydrodynamic phase is completed, particles are sampled from the freeze-out hypersurface and fed into UrQMD.  For each collision event, UrQMD yields a list of particles which were emitted by that collision, together with their momentum space coordinates, after all rescattering has finished.  The particle lists generated by UrQMD are then used as input to the PH pipeline described in Sec.~\ref{sec:overview_ph}.  Since our focus in this work is on PH observables, we do not explore more standard nuclear collision observables here, such as those which have already been extensively studied and compared with data elsewhere \cite{Moreland:2018gsh, Bernhard:2019bmu}.

\subsection{Centrality classes and background construction}
In order to study the centrality dependence of PH observables, we divide the ensemble of events into ten deciles.  The multiplicity distribution is shown versus centrality class in Fig. \ref{fig:centrality}. We exclude collisions with output multiplicity less than fifty particles, as the PH statistics are rendered highly unstable for very small point clouds. Smaller event multiplicities can be probed by considering a sufficiently large number of events, a task we defer to future work.

Once the simulated nuclear collision data has been generated, we construct a background against which to compare the PH observables extracted from individual events.  The way we do this is similar in spirit to the usual `mixed event' approach employed in experimental analyses (e.g., Ref.~\cite{CMS:2013jlh}): we combine all simulated events in the same centrality class into a single large, uncorrelated event, where each event has been rotated by a different random angle $\delta\phi \in [0,2\pi]$.  We then sample uniformly from this combined event to obtain an event in the same centrality class as the original events.  The original events are referred to as `signal' events, while the mixed events are referred to as `background' events.  This procedure thus provides a reference against which to test the significance of our PH observables.

\subsection{PH analysis}

Each signal or background event includes a list of discrete particles with momentum-space coordinates. Each event is individually supplied as input to our PH pipeline, so that our analysis is carried out on an event-by-event basis.  This yields an ensemble of signal PH observables and another ensemble of background PH observables, where the latter are used to establish a baseline for the former.

For the PH calculations themselves we use the open-source computational package \texttt{giotto-ph} \cite{perez2021giotto}. The \texttt{treelib} package is used for generating the dendrograms.  With these tools we construct and analyze the four different PH observables discussed above: (i), fractal dimension; (ii), Betti curves; (iii), cluster entropy; and (iv), local clustering statistics. Once the observables are constructed for both sets of events, either the ratio or the difference between the signal and background is taken, depending on the specific observable under consideration (as discussed below).

\section{Results}\label{sec:results}
In this section we describe our results obtained by applying PH to point clouds generated from Pb+Pb collisions and compare our PH observables in signal events to those generated from background events. We employ the novel statistical summaries outlined in Sec. \ref{sec:overview_ph} and present our results within each centrality class unless otherwise specified.

For several observables we compute the difference between the signal and background events rather than the ratio. Our reasoning is that several topological summaries describe the number or magnitude of homological features as a function of filtration, and therefore the difference in topology is functionally more appropriate than the ratio. This is in contrast to $n$-point correlation functions, wherein ``dividing out" the background is a more natural procedure \cite{CMS:2013jlh}.\par  
For our PH pipeline we employ a sublevel set filtration of the functional $\ell(v) = \left(\sum_{t\in \Delta(v)}\text{Area}(t)\right)^{1/2}$; here $v$ is a vertex (point) in the Delaunay triangulation, $\Delta(v)$ is the set of triangles adjacent to $v$, and $\text{Area}(t)$ denotes the area of the triangle $t$ in the $(\phi,y)$ plane. As discussed in Appendix \ref{sec:app_dtfe}, this sublevel set filtration is equivalent to a superlevel set filtration of a density functional; the square-root ensures units of (angular) distance. Put more plainly, filtering from small to large values of $\ell$ is equivalent to (up to a monotonic map) filtering from large to small density. \par 
To assess the effect of non-topological density fluctuations, we also consider some of the observables under a modified filtration $\tilde{\ell}:= \ell /\langle\ell\rangle$, where $\langle \ell \rangle $ denotes the mean value of $\ell$ \textit{within an event}. This filtration modification is performed for all events in a centrality class prior to computing an observable, and we explicitly note both in the text and in figures which filtration we use.  \par
Finally, for the fractal dimension calculations we omit any infinitely long-lived homological features. In 0D this omission corresponds to the largest connected component, while in 1D we omit the topological loop indicative of the topology of the cylinder. 
\par

\subsection{Persistence Diagrams}
We begin our analysis by examining how the persistence diagrams from the signal collisions differ from those of the background events. For each event type (signal or background) we aggregate the persistence diagrams within each centrality class and compute a count-normalized 2D histogram. We then compute the difference in histograms between the signal events and the background events, the results of which for homological dimension zero and one are depicted in Fig. \ref{fig:pers_diag_0} and Fig. \ref{fig:pers_diag_1}, respectively. Each subplot for centrality classes $0-10\%$ through $70-80\%$ depicts by color where the persistent homology of the two event types substantially differ: red regions indicate where the signal events have a stronger concentration of persistent topological features, while the blue regions indicate where the signal events have fewer persistent features with respect to the background. \par 
We first note that there is a strong tendency for the signal collisions to have a concentration of persistence features at earlier filtration values, as most easily seen in Fig. \ref{fig:pers_diag_0}(c)-(e). As we travel up the diagonal, we notice that the concentration of signal persistence features give way to a suppression of persistent features relative to the background (the blue "stripe" running perpendicular to the diagonal). This stripe is less prominent for $50-60\% $ and higher centrality classes.  This is consistent with the presence of stronger elliptic flow $v_2$ in mid-central collisions than in central collisions \cite{ALICE:2010suc}, which produces more points in-plane than out-of-plane and thus leads to a more rapid formation of structure as a function of filtration than an event with vanishing $v_2$.

\begin{figure}
    \centering
    \begin{subfigure}{\linewidth}
    \centering
    \includegraphics[]{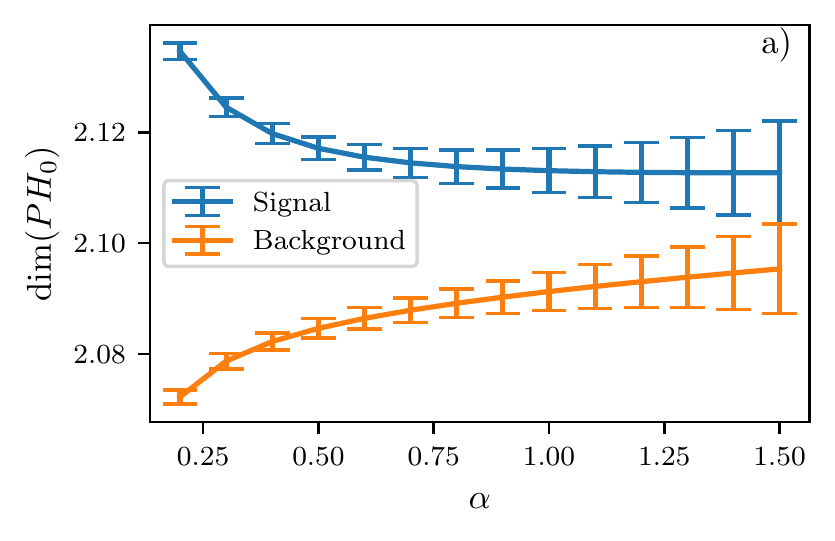}
    \end{subfigure}
    \begin{subfigure}{\linewidth}
    \centering
    \includegraphics[]{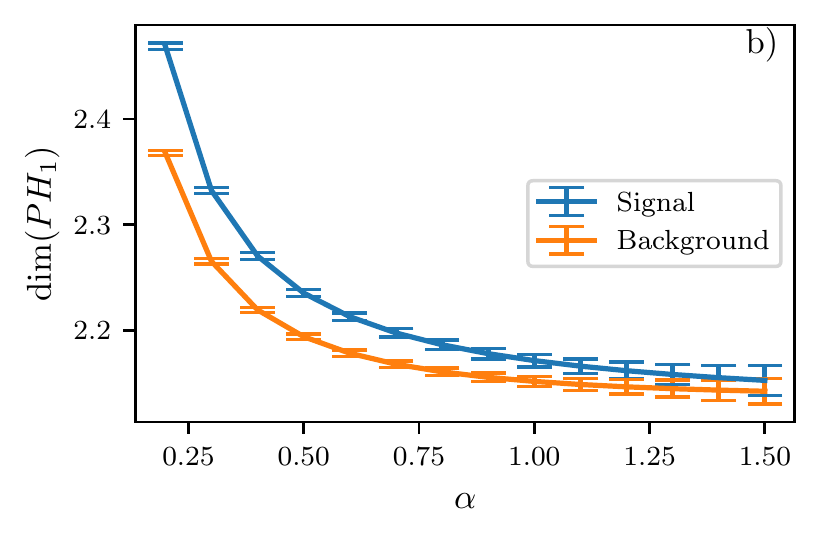}
    \end{subfigure}
    \caption{a) Fractal dimension $\dim(PH_{0})$, defined in Eq. \ref{eq:beta}) as a function of $\alpha$. a) Fractal dimension for 0D homology. b) Difference in 0D fractal dimension between signal and background. c) Fractal dimension for 1D homology. d). Difference in 1D fractal dimension between signal and background. }
    \label{fig:f_dim_0}
\end{figure}

Similar to the 0D case, we also compare the 1D persistence diagrams  in Fig. \ref{fig:pers_diag_1}. We see again the striping behavior (predominance of the signal lifetimes early in the filtration, followed by a suppression relative to the background). The 1D PDs display a different overall shape from the 0D PDs: the 0D PDs show a broader distribution of death values early in the filtration, while the 1D PDs exhibit a distribution of death value that broadens later in the filtration. This implies that loops born later in the filtration persist over a larger range of filtration values. Note as well that, due to our density-based filtration, the 1D PD is effectively measuring the propensity and relative scale of high-density regions surrounding low-density regions. These fluctuations can be interpreted in terms of local curvature and quantified through fractal dimensions, as we explore next.

\subsection{Fractal dimensions}

As noted above in Sec. \ref{sec:observables}, the fractal dimension can serve as a measure of fractality and local curvature. Recall that the fractal dimension effectively measures how the $p$-norm (here we use $\alpha$ instead of $p$) of the PD lifetimes scales with the multiplicity of the underlying point cloud. A small $\alpha$ probes local curvature, while large $\alpha$ probes more global structure.\par  
In Fig. \ref{fig:f_dim_0}a) we depict the fractal dimension of the 0D homology as a function of $\alpha$, both for the signal and background events. The standard error is estimated from a linear regression of the slope $\beta$ and then propagated through to the fractal dimension. Fig. \ref{fig:f_dim_0}b) shows the difference in fractal dimensions between signal and background events. We first note that, for small $\alpha$, the difference in fractal dimension between signal and background is quite substantial, indicating that the persistent homology identifies a higher degree of clustering and fractality in the signal collisions for 0D. Given that the ambient space is effectively a closed cylinder, it is perhaps not surprising that the fractal dimensions are $\sim 2$. For larger values of $\alpha$ both the signal and background collisions steadily converge and becomes statistically hard to distinguish. Given that our analysis is for mid-rapidity observables $|y|\leq 2$, this lack of distinction follows from the limited volume of our bounded cylinder.\par 
In Fig. \ref{fig:f_dim_0}c-d) we show the fractal dimension for the 1D homology and the difference in fractal dimension between the signal and background events. Curiously, for one-dimensional PH both the signal and background events have the same monotonic decrease in dimension, though the fractal dimension differences widen as a function of $\alpha$. Moreover, in both 0D and 1D the fractal dimension for the signal events is higher than the background. A simple explanation is that the higher degree of clustering in the signal events tends to form shorted-lived loops. As the $\alpha$-norms for $\alpha<1$ emphasize small features, this implies the fractal dimension for 1D is larger for the signal events, though the gap appears to close for sufficiently large $\alpha$.

\subsection{Betti curves}
While the fractal dimension yields important distinctions between the signal and background events, the fractal dimension is insensitive to when in the filtration homological features are most prominent. The Betti curve, described in Sec. \ref{sec:observables}, gives more precise insight into the distribution of homological features as a function of filtration.  

\subsubsection{Mean of $\beta_i(\ell)$}

\begin{figure*}
\centering
    \begin{subfigure}{.247\textwidth}
    \centering 
    \includegraphics[width=\textwidth]{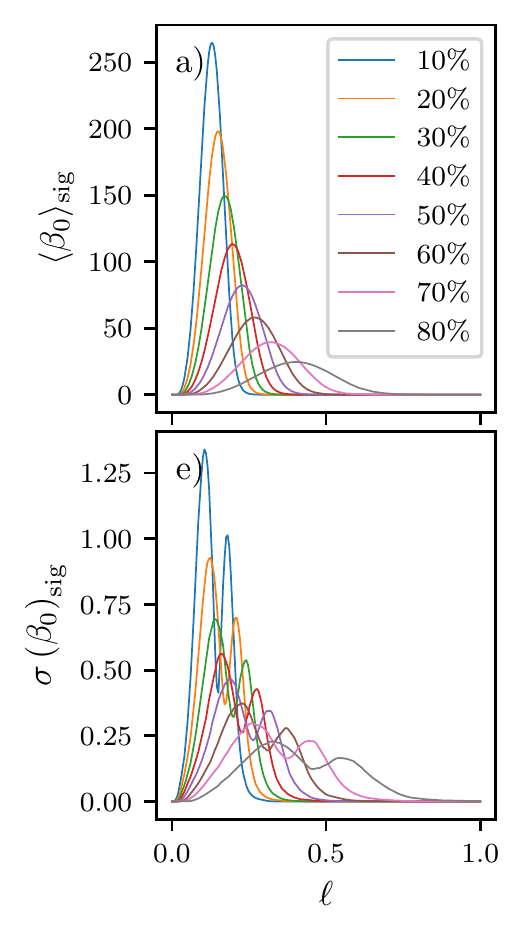}
    \end{subfigure}%
    \begin{subfigure}{.24\textwidth}
    \centering 
    \includegraphics[width=\textwidth]{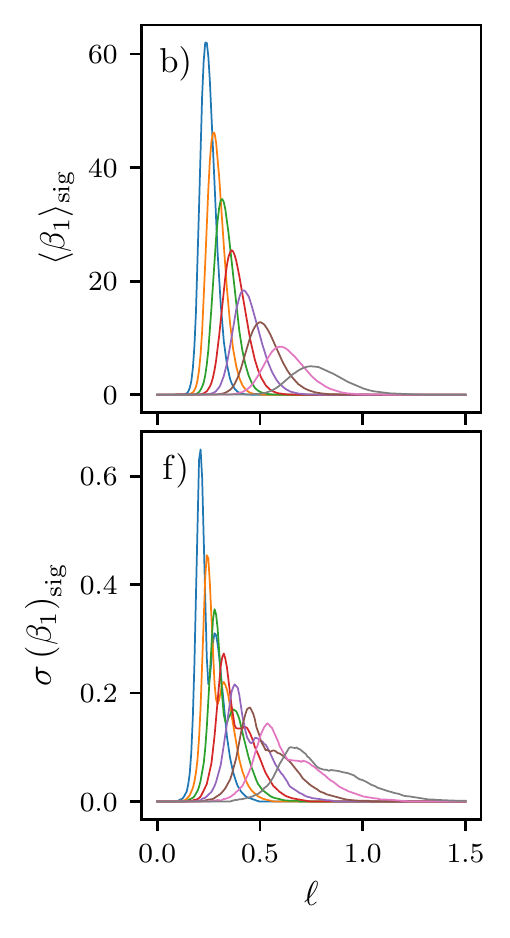}
    \end{subfigure}%
    \begin{subfigure}{.24\textwidth}
    \centering 
    \includegraphics[width=\textwidth]{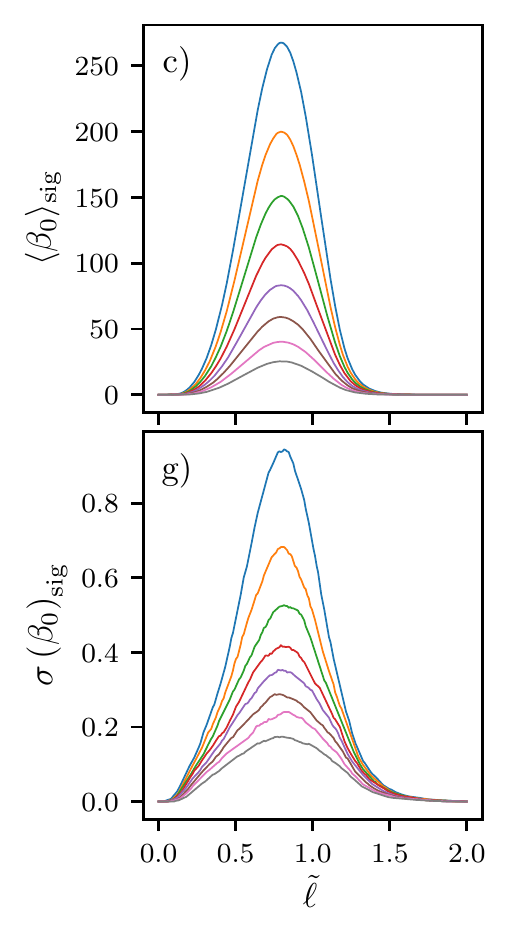}
    \end{subfigure}
    \begin{subfigure}{.24\textwidth}
    \centering 
    \includegraphics[width=\textwidth]{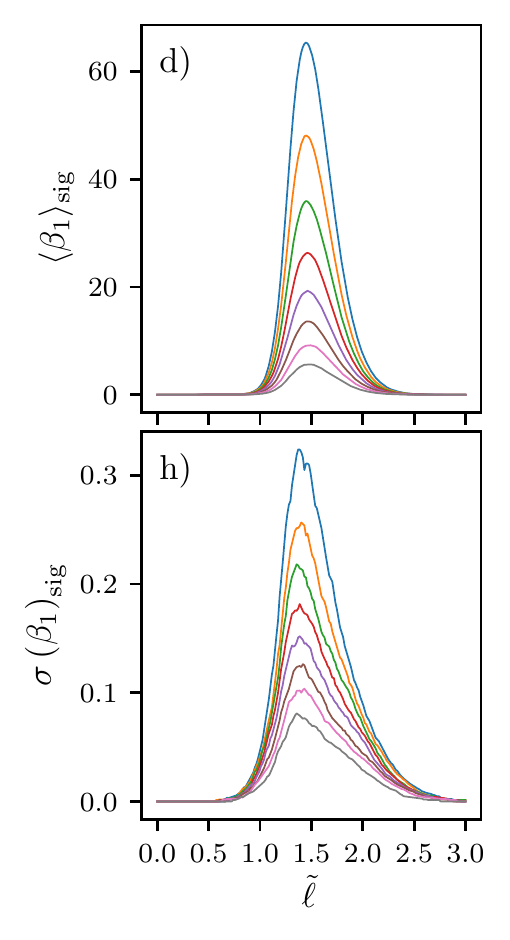}
    \end{subfigure}%

    \caption{a) Mean $\beta_{0}$, calculated by averaging within a centrality class while holding a set of sampling filtration values fixed. b) Mean $\beta_{0}$ in the $\tilde{\ell}$ filtration. c) Standard error of $\beta_{0}(\ell)$. d) Standard error of $\beta_{0}(\tilde{\ell})$.}
    \label{fig:betti_curves}
\end{figure*}

In Fig. \ref{fig:betti_curves}a-h) we show the mean and standard error of the Betti curves $\beta_{i}(\ell)$ and $\beta_{i}(\tilde{\ell})$ for signal collisions in each centrality class.

\par

To construct the mean and standard error of the signal Betti curves, we compute the average (or standard error) over all events while holding the filtration value fixed. Repeating this process for a large number of filtration values yields Fig. \ref{fig:betti_curves}. The mean $\beta_{i}$ curves both start at zero and end at one. For the $\beta_{0}$ curve, we begin the filtration with no clusters and end with the cluster representing the entire point cloud. For the $\beta_{1}$ curve, we begin with no loops and end with no loops apart from the topological loop represented by the closed cylinder. However, this loop has infinite lifetime and is therefore ignored in the persistence diagram and fractal dimension calculations. Due to the DTFE construction and our filtration definition, clusters can appear at any point during the filtration.

Fig. \ref{fig:betti_curves}a shows that each centrality class possesses a single peak, located at a value of $\ell$ which increases with decreasing multiplicity. This has a straightforward interpretation: events with large multiplicity have a smaller average inter-particle spacing than events with small multiplicity, which distribute a smaller number of particles over the same region in momentum space. Consequently, large multiplicity events will establish connections at smaller values of the filtration parameter, and conversely for small multiplicities. Similarly, the merging of separate clusters (and eventual reduction of $\beta_0$ to zero) also proceeds more rapidly with $\ell$ at high multiplicity than low multiplicity. As a result, the peak is reached first by the most central collisions and last by the most peripheral collisions. We note in passing that the Betti curve of each centrality class in Fig. \ref{fig:betti_curves}a is very well described by a Weibull distribution, which has been shown elsewhere to reproduce well the distribution of cluster sizes and nearest neighbor distributions in random point clouds \cite{RevModPhys.15.1}. Although we do not further explore this issue here, we speculate that measuring \textit{deviations} of $\beta_0$-curves from a Weibull distribution may provide a useful way of quantifying non-trivial correlations in point clouds in general. We defer a careful discussion of this possibility to future work.\par 
In Fig. \ref{fig:betti_curves}c we depict $\beta_{0}(\tilde{\ell})$; note that under this rescaling the peaks of the Betti curve all roughly coincide. This supports the conclusion that the rate of cluster formation and inter-particle spacing distributions should depend strongly on multiplicity and therefore the density; thus, rescaling the filtration removes some, but not all of this effect.\par 
 Fig. \ref{fig:betti_curves}b shows the mean $\beta_{1}(\ell)$ curve, wherein we observe similar behavior as the $\beta_{0}(\ell)$ curve: the maximal number of non-homologous loops existing at any point in the filtration steadily decreased with multiplicity, while the location of the peak increases in filtration value as a function of centrality class. Rescaling to $\tilde{\ell}$ yields Fig. \ref{fig:betti_curves}d; again we see the peak $\beta_{1}(\tilde{\ell})$ all roughly coincide across centrality classes.

\subsubsection{Standard error of $\beta_i(\ell)$}

In Fig. \ref{fig:betti_curves}e we show the standard deviation of the Betti curve $\beta_{0}(\ell)$ (i.e., the fluctuations about the mean Betti curve) for each centrality class. The $\sigma(\beta_0)$ in each centrality class exhibits two distinctive peaks which are most prominent in the most central collisions. We note also that the second peak is typically slightly smaller than the first. This can again be straightforwardly understood in terms of event-by-event fluctuations in the scale and location parameters of the underlying Weibull distribution extracted from a single nuclear collision. The slightly smaller second peak is then a consequence of the resulting fluctuations in the shallower slope as $\langle \beta_0 \rangle$ descends from its peak value. Both properties of the complete $\beta_0$-distribution shown in Fig.~\ref{fig:betti_curves} are thus consistent with fluctuations of the average density within a single centrality class. As a confirmation of this analysis, the bimodality disappears under the rescaled $\tilde{\ell}$ filtration, as shown in Fig. \ref{fig:betti_curves}g. \par 
We show the standard error for the $\beta_{1}(\ell), \, \beta_{1}(\tilde{\ell})$ curve in Fig. \ref{fig:betti_curves}f and h. Quite similar to the $\beta_{0}(\ell)$ standard error, we observe a bimodality that disappears under the rescaling $\ell \to \tilde{\ell}$.

\begin{figure*}
    \centering
    \begin{subfigure}{.47\textwidth}
    \centering
    \includegraphics[width=\textwidth]{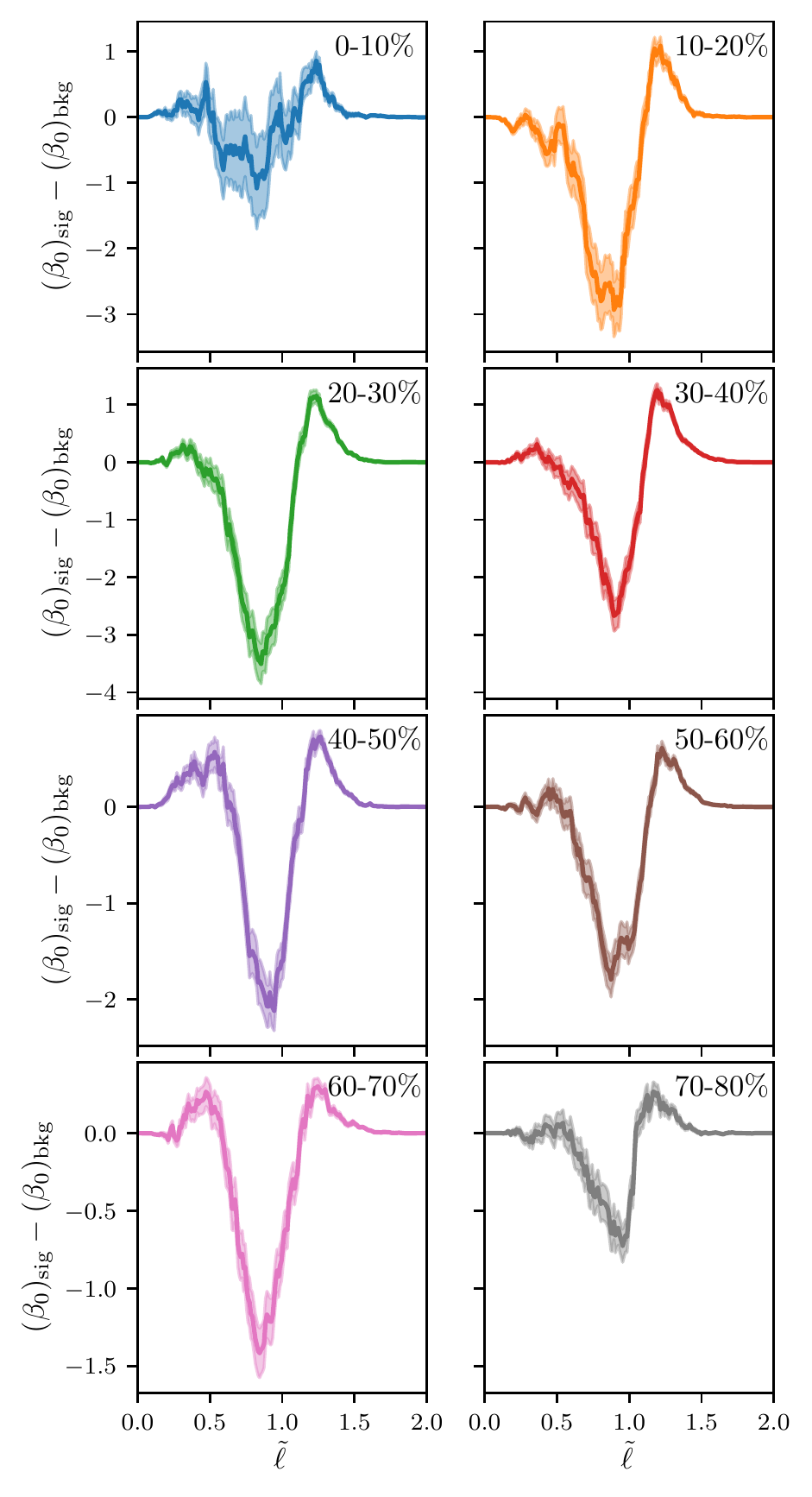}
    \end{subfigure}
    \begin{subfigure}{.48\textwidth}
    \centering
    \includegraphics[width=\textwidth,trim={0cm .3cm 0cm 0.25cm},clip]{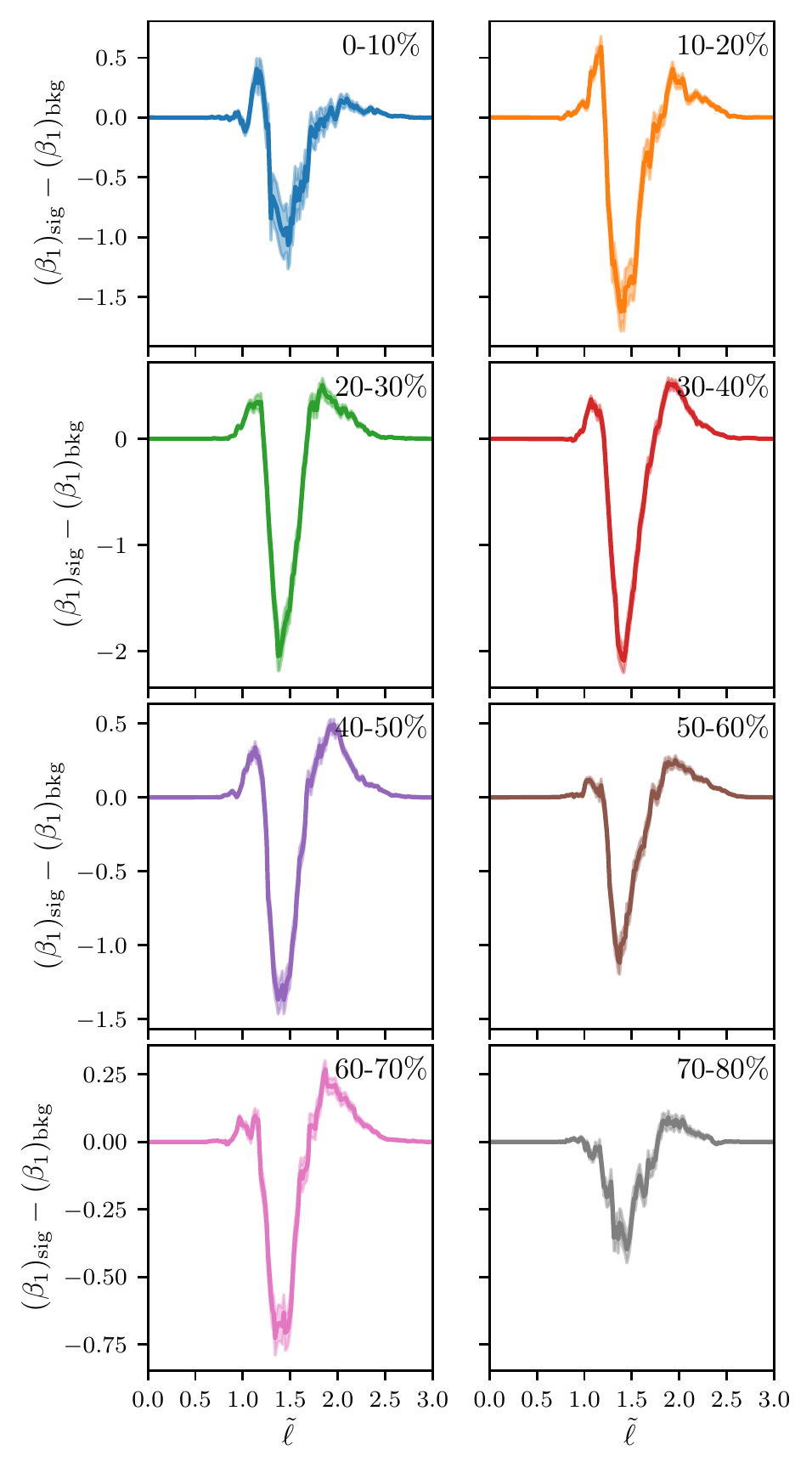}
    \end{subfigure}
    \caption{(left) Difference in $ \beta_{0}(\tilde{\ell})$ between the signal and background for each centrality class. (right) Difference in $ \beta_{1}(\tilde{\ell})$ for each centrality class.}
    \label{fig:betti_curve_diff_0}
\end{figure*}

\begin{figure*}
    \centering
    \includegraphics[]{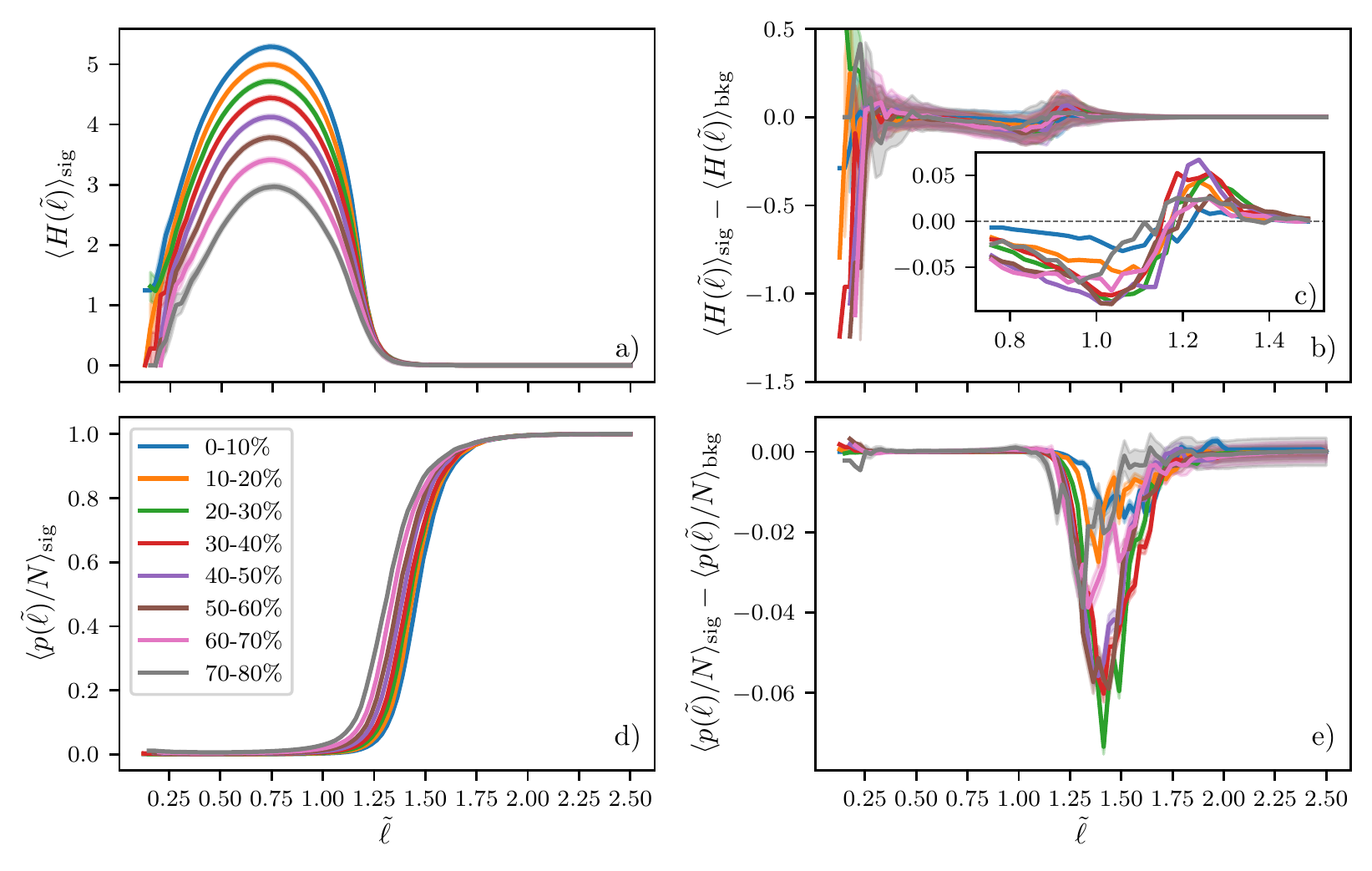}
    \caption{Cluster statistics as a function of centrality class and filtration. Each plot is computed by averaging over events in a centrality class with a set of sampling filtration values fixed. a) Average cluster entropy $H(\tilde{\ell})$ for the signal events. b) Difference in cluster entropy between the signal and background events with enhanced resolution in the inset c) (the uncertainty bands have been removed from the inset to enhance visibility). 
    d) Mean number of points per cluster, dividing by the multiplicity. e) Difference in mean number of points per cluster (dividing by the multiplicity) between the signal and background events.}
    \label{fig:hier_entropy}
\end{figure*}

\subsubsection{$\beta_i(\ell)$ difference}

In Fig. \ref{fig:betti_curve_diff_0} we show the difference between the signal and background events in mean $\beta_{i}(\tilde{\ell})$: the left panel shows $\beta_{0}(\tilde{\ell})$ while the right shows $\beta_{1}(\tilde{\ell})$. \par 
Beginning with $\beta_{0}$, we note for all centrality classes common behavior: a small peak in the signal $\beta_{0}$, followed by a large dip wherein the signal $\beta_{0}$ is lower than the background $\beta_{0}$, followed by a final larger peak in the signal $\beta_{0}$ relative to the background. \par 
The initial uptick in $\beta_{0}$ and large dip at $\tilde{\ell} \approx 1$ is of course consistent with our expectation, already reflected in Figs.~\ref{fig:pers_diag_0} and \ref{fig:pers_diag_1}, to have more clustering earlier in the filtration, and thus a lower number of clusters and a smaller Betti number than the corresponding background. However, the tendency to have more clustering at the beginning of the filtration appears to drop off for larger centrality classes (see the $50-60\%$ and $60-70\%$ centrality classes). This characteristic rise-and-dip pattern observed in the mid-central collisions implies an initial enhancement of clustering in the signal events as compared with the background which is a direct consequence of enhanced local clustering resulting from collective, anisotropic flow. The final peak in $\beta_{0}(\tilde{\ell})$ implies that, late in the filtration, there is a resurgence of clusters in the signal events relative to the background. Since late filtration corresponds to low-density regions of the point cloud, the signal events have a preference for stronger fluctuations in density within low-density regions, particularly in low centrality classes. 
\par 
For the $\beta_{1}(\tilde{\ell})$ difference we observe a similar rise-and-dip behavior, though for the $0-10\%$ and $10-20\%$ centrality classes the initial peak is stronger than the final peak. Furthermore, the locations of the peaks and dips occur considerably later in the $\tilde{\ell}$ filtration than the corresponding features in the $\beta_{0}$ curves.  This is consistent with the onset of loop formation occurring somewhat later than the beginning of cluster formation, relative to the background events. 

\subsection{Cluster Entropy}
As noted above, the Betti curve is indifferent to the relative distribution of points between clusters at each level of the filtration. Since our filtration amounts to a hierarchical clustering scheme, we can access the number of points per cluster and calculate a ``cluster entropy", defined in Sec. \ref{sec:observables}. To reiterate, the cluster entropy $H(\tilde{\ell})$ is given by \begin{align}
    H(\tilde{\ell}) = -\sum_{C_{i} \in \mathcal{C}(\tilde{\ell})} p_{i}(\tilde{\ell}) \log p_{i}(\tilde{\ell}),
\end{align} where $i \in \mathcal{C}(\tilde{\ell})$ is the set of clusters at value $\tilde{\ell}$ and $p_{i}(\tilde{\ell}) = |C_{i}|/n(\tilde{\ell})$, $n(\tilde{\ell})$ being the number of points that exist at value $\tilde{\ell}$ (recall that, by our DTFE, points enter during the filtration and thus do not exist at the beginning of the filtration). \par 
In what follows we also compute the mean number of points per cluster $p(\tilde{\ell}) = \frac{1}{|\mathcal{C}|}\sum_{C_{i} \in \mathcal{C}} p_{i}(\tilde{\ell})$. To ensure fair comparison of events within the same centrality class but different multiplicity, we divide by the multiplicity $N$ and compute $p(\tilde{\ell})/N$.
\par 
In Fig. \ref{fig:hier_entropy}a we show the mean cluster entropy $H(\tilde{\ell})$ for the signal events as a function of filtration and centrality classes. Note for each centrality class the curve begins at the formation of the first cluster, as the entropy is undefined prior to this point. We observe across all centrality class a rise in the cluster entropy to a peak that scales in magnitude with the average multiplicity of the centrality class, followed by a steep descent to a vanishing cluster entropy which coincides with the merging of all points into one connected component. \par 
In Fig. \ref{fig:hier_entropy}b) we show the difference in cluster entropy between the signal and background events. Strong fluctuations in the cluster entropy for small $\tilde{\ell}$ evolve into a consistent supression in signal cluster entropy around $\tilde{\ell} \approx 1.0$, followed by an increase at $\tilde{\ell} \approx 1.25$ before the signal and background cluster entropies converge to a common value of zero. \par 
In Fig. \ref{fig:hier_entropy}c we depict the $p(\tilde{\ell})/N$ for the signal events and plot the difference in $p(\tilde{\ell})/N$ between the signal and background in Fig. \ref{fig:hier_entropy}d. Note that for all centrality classes save the $70-80\%$ class the difference in $p(\tilde{\ell})/N$ is positive early in the filtration, indicating signal events have more points per cluster. This reflects our expectation that, early in the filtration, more local clustering implies more clusters to start relative to the background. At $\tilde{\ell} \approx 1.25$ the mean points per cluster is strongly supressed for the signal events relative to the background. Note that by this time in the filtration the cluster entropies of the signal and background have largely converged. This sharp reduction in mean points per cluster coincides with the uptick in $\beta_{0}(\tilde{\ell})$ in Fig. \ref{fig:betti_curve_diff_0} around $1.0\le \tilde{\ell} \le 1.5 $, which indicates the signal point clouds have a late-filtration increase of clusters with small multiplicities. These small clusters drive down $p(\tilde{\ell})$. Due to the small size of these new clusters, the cluster entropy is relatively unchanged, as the largest size cluster dominates.

\subsection{Local Clustering Statistics}
Each of the observables so far explored are completely insensitive to the relative spatial degrees of freedom in the point cloud; i.e., the actual locations of the hadrons in $(\phi,y)$ space. Put differently, none of the observables would change under a rotational or Lorentz boost along the beam axis. This is unfortunate, as one might expect some dynamical processes to introduce anisotropies, the spatial statistics of which would be of interest. This is particularly true for hydrodynamical flow, the effects of which we have seen above are implicit in a number of the observables discussed previously.

As noted in Sec. \ref{sec:observables}, however, one benefit of the PH pipeline is a dendrogram which reflects local degrees of freedom in the system and captures which clusters merge when in the filtration. Using the local clustering statistic introduced in Sec. \ref{sec:observables}, we calculated, for each 0D PD, the $p$-norm of the $\bm{t}_{i}$ for each hadron. To recall, $\bm{t}_{i}$ for hadron indexed $i$ has as components the lengths of branches between merges of clusters containing $i$ (c.f. Fig. \ref{fig:merge_tree}). 
Our process for computing the local clustering statistics is as follows. First, we computed for each dendrogram the $p$-norm $||t_{i}||^{p}$ as a function of leaf $i$, which we define as $f_{p}(\phi_{i})$. Second, we determined the mean $p$-norm within a centrality class and within equal-sized bins (width $.01$ radians) of the azimuthal angle $\phi$. Note that we integrate over the rapidity range $|y|\le 2$, by our DTFE construction. We denote the average $p$-norm as $\langle f_{p}(\phi)\rangle_{X}$, where $X$ denotes averaging over either the signal (sig) or background (bkg) events and $\phi$ now denotes a bin. Third, we took the ratio $g(\phi):=\langle f_{p}(\phi)\rangle_{sig}/\langle f_{p}(\phi)\rangle_{bkg}$. Finally, we extracted the $v_{2}$ Fourier coefficient of $g(\phi)$, normalized by the average of $g(\phi)$; in particular, we follow standard practice \cite{Luzum:2013yya} and define the complex quantity 
\begin{align}
    V_2 \equiv v_2 e^{2 i \Psi_2}  = \frac{\int^{2\pi}_0 g(\phi)e^{2i\phi}d\phi}{\int^{2\pi}_0 g(\phi)d\phi}. \label{pnorm_v2}
\end{align}

We emphasize that, despite using notation similar to that which is normally used in nuclear collision phenomenology, the quantity Eq.~\eqref{pnorm_v2} should \textit{not} be confused with the usual measure of elliptic flow, which is computed in a completely different way \cite{Heinz:2013th}.

\begin{figure}
    \centering
    \includegraphics[]{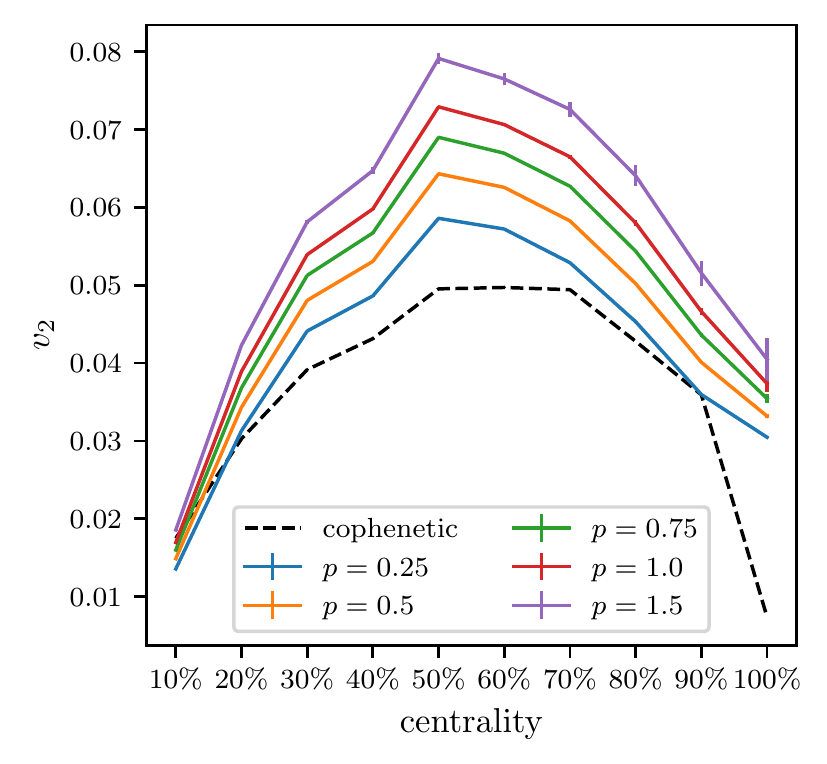}
    \caption{Extracted $v_{2}$ coefficient of the average $p$-norm local clustering statistic, shown as a function of centrality class and $p$. The dashed line represents the $v_{2}$ coefficient extracted from the cophenetic distance function show in Fig. \ref{fig:coph}.}
    \label{fig:local_v2}
\end{figure}

We evaluated Eq.~\eqref{pnorm_v2} for $p = \{0.25,0.5,0.75,1.0,1.5\}$ and within each centrality class. Fig. \ref{fig:local_v2} shows the extracted $v_{2}$ coefficient as a function of centrality class and $p$-norm. The distinctive peak in the flow magnitude $v_{2}$ in mid-central collisions is characteristic of the geometry-driven flow anisotropy observed in nucleus-nucleus collisions \cite{STAR:2004jwm, ALICE:2016ccg}.

\begin{figure*}
    \centering
    \includegraphics[]{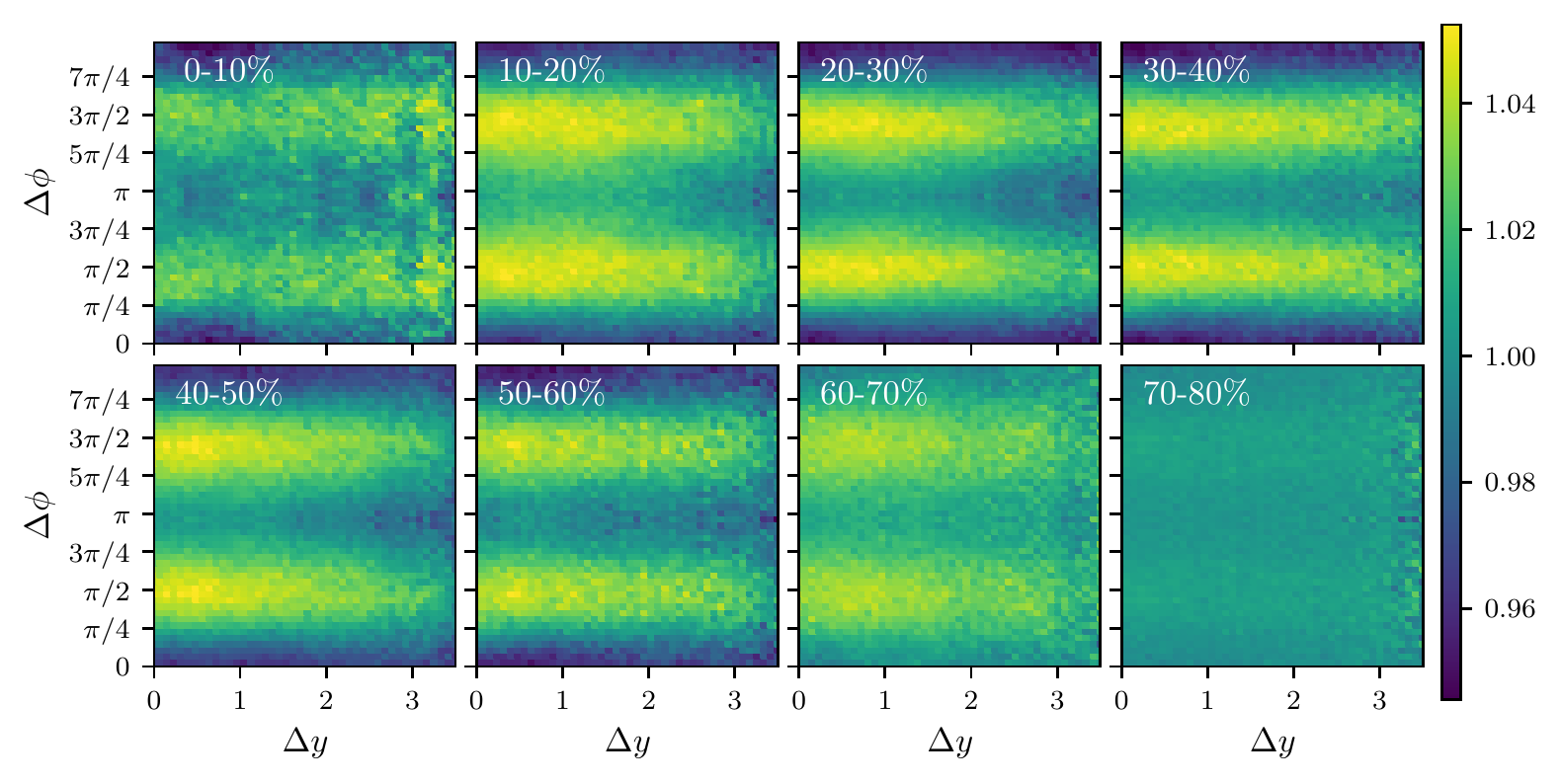}
    \caption{The ratio of cophenetic correlation functions $d_{c}^{sig}(\Delta \phi, \Delta y)/d_{c}^{bkg}(\Delta \phi,\Delta y)$ for different centrality classes. We note the long-range (in rapidity) enhancement which exhibits the oscillation characteristic of elliptic flow.}
    \label{fig:coph}
\end{figure*}
This demonstrates that anisotropic flow can indeed be accessed and quantified using PH. Note that for illustrative purposes we have ignored complications of our procedure relating to the estimation of the elliptic flow plane \cite{CMS:2012zex}, which we take in this work to coincide with the positive $x$-axis in the transverse plane. In principle, sensitivity of flow observables to the $n$-th flow plane angle $\Psi_n$ can be significant and is typically avoided by working instead with multi-particle correlation functions, which by construction do not depend on $\Psi_n$ \cite{Heinz:2013th}.

A thorough generalization of the analysis we present here to one which is similarly insensitive to $\Psi_n$ will be deferred to future work.  However, we briefly consider one possible way of doing this which involves introducing a PH-based notion of separation between pairs of hadrons.  This notion is provided by the cophenetic distance, which we discuss next.

\subsection{Cophenetic distance}

As noted above, the dendrogram encodes both the clustering information (through the heights at which branches merge) as well as positional information (by looking at the positional coordinates of the leaves). One natural statistic that couples the physical degrees of freedom to the clustering statistics is the cophenetic distance. The cophenetic distance $d_{c}(i,j)$ between two points $i,j$ is the height in the dendrogram (value of the filtration) at which the corresponding leaves merge into a single cluster \cite{sokal1962comparison}. 
 
\par 
This distance function can be naturally extended to a correlation functional $d_{c}(\Delta \phi,\Delta y):= \langle \langle d_{c}(i,j) \rangle \rangle $ such that points $i,j$ have separation $(\Delta \phi,\Delta y)$, and $\langle \langle \rangle \rangle$ denotes averaging over dendrograms within a centrality class. This \textit{cophenetic correlation function} is readily generalizable to $n$-th order correlation functions (e.g., the cophenetic distance between three points, etc.).\par 
In Fig. \ref{fig:coph} we depict the ratio of cophenetic distance functions $d_{c}^{sig}(\Delta \phi, \Delta y)/d_{c}^{bkg}(\Delta \phi,\Delta y)$, where the superscripts indicate the type of event. 
As in the case of the local $p$-norm clustering statistics, we extracted a corresponding $v_{2}$ Fourier coefficient for the correlation distance function.  Defining
\begin{align}
\tilde{g}(\Delta\phi) \equiv \int\displaylimits^{\Delta y_\text{max}}_{-\Delta y_\text{max}} d\Delta y\, \frac{d_{c}^{sig}(\Delta \phi, \Delta y)}{d_{c}^{bkg}(\Delta \phi,\Delta y)},
\end{align}
and setting $\Delta y_\text{max} = 3$, we compute the cophenetic $v_2$ coefficient according to
\begin{align}
    v^\text{cophenetic}_2  = \left(\frac{\int^{2\pi}_0 \tilde{g}(\Delta\phi)e^{2i\Delta\phi}d\Delta\phi}{\int^{2\pi}_0 \tilde{g}(\Delta\phi)d\Delta\phi}\right)^{1/2}
\end{align}
The result is plotted in Fig. \ref{fig:local_v2}. The cophenetic $v_{2}$ shows good agreement with the $v_{2}$ coefficient from the local clustering.

A large cophenetic distance between two particles implies a longer time for them to merge into a single cluster, so that sparsely populated regions lead to an enhanced cophenetic distance (relative to the background); conversely, densely populated regions produce suppression of the cophenetic distance in those region.  Thus the elliptic flow signal emerges here (with peaks at $\Delta \phi \approx \pi/2$ and $3\pi/2$ and valleys at $\Delta \phi \approx 0$ and $\pi$), qualitatively consistent with that obtained by other two-particle methods \cite{STAR:2004jwm, CMS:2012zex, ALICE:2016ccg}.  \par

\section{Conclusions}

In this work we have demonstrated how to use persistent homology to probe correlational and topological structures in the particle distributions produced by nuclear collisions. Our primary aim has been to advance the notion of a point cloud as a useful and flexible perspective for characterizing the properties of these particle distributions. We have utilized a density-based filtration of ensembles of point clouds to identify large and small-scale topological structure, and introduced several new tools and observables to probe aspects of nuclear collision phenomenology. Most importantly, we have augmented existing tools from topological data analysis to incorporate spatial degrees of freedom, in an effort specifically to quantify anisotropies indicative of hydrodynamical flow. While these topologically-minded observables have close, intuitive connections to traditional correlational measures, it remains the subject of future work to formalize the relationships between PH observables and standard statistical measures. In this vein, we briefly note several directions in which our analysis could be improved.\par

Applications of PH to nuclear collisions have been largely unexplored save for a small collection of works \cite{li2020jet}. One fundamental challenge of PH is cleanly relating PH observables to traditional correlational measures like $n$-point connected correlation functionals. The density-based filtration used here has strong connections to Minkowski functionals, Euler characteristics, and integrals of connected correlation functions \cite{Schmalzing_1997,Schmalzing_Buchert_1997}, but a ``dictionary" mapping from PH to standard statistical measures has (to our knowledge) yet to be constructed.

Our PH observables are also closely tied to recent work on energy flow polynomials, where correlational structures are tied to functionals of the angular degrees of freedom in the resultant point cloud \cite{komiske2018energy}.  Another further direction of pursuit is defining robust observables that better incorporate the positional degrees of freedom, like the local clustering statistic given here. One could imagine statistics like cluster correlation functions being generalized to PH observables, and we intend in future work to explore this further.

Finally, PH is not limited solely to point clouds and could profitably be applied to other aspects of nuclear collisions. For instance, PH has been successfully applied to discretized scalar fields (e.g., on a hyper-lattice) through the use of \textit{cubical} complexes \cite{Kramar_Levanger_Tithof_Suri_Xu_Paul_Schatz_Mischaikow_2016,kaji2020cubical}. In the hydrodynamical phase of the evolution of the QGP one could leverage PH at each phase of the flow to form a sequence of topological fingerprint ``snapshots" of the dynamical system. The evolution of these snapshots as a time series can be further explored by leveraging distance measures between PDs \cite{Kramar_Levanger_Tithof_Suri_Xu_Paul_Schatz_Mischaikow_2016}. A recent exploration of using PH to probe Rayleigh convection leveraged these tools to identify Lyapunov exponents and critical behavior \cite{Kramar_Levanger_Tithof_Suri_Xu_Paul_Schatz_Mischaikow_2016}. A similar pipeline could be used to probe the emergence of turbulent or critical behavior of QGP.

\bigskip

\section*{Acknowledgements }
The authors would like to thank Jorge Noronha and Jacquelyn Noronha-Hostler for insightful discussions on potential applications of persistent homology to nuclear collisions. C. P. was supported by the U.S. DOE Grant No. DE-SC0020633. G.H. acknowledges support from the Department of Energy grant DOE DESC0020165. T. D. acknowledges support from the ICASU Graduate Fellowship. This work made use of the Illinois Campus Cluster, a computing resource that is operated by the Illinois Campus Cluster Program (ICCP) in conjunction with the National Center for Supercomputing Applications (NCSA) and which is supported by funds from the University of Illinois at Urbana-Champaign.

\bibliography{references,not_inspire, inspire}

\appendix
\section{Details of DTFE}\label{sec:app_dtfe}
In this appendix we more fully describe the DTFE, outline the correspondence between superlevel and sublevel set filtrations, and document our procedure for implementing the DTFE in the context of nuclear collisions.\par

Given the Delaunay triangulation of a point cloud in an ambient space, let $v$ denote a vertex in the triangulation, and define $\Delta_{v}$ as the set of $n$-simplices adjacent to $v$. We define a functional \begin{align}\label{eq:dtfe}
    f_{p}(v) = \left(\sum_{t \in \Delta(v)}V(t)\right)^{p},
\end{align} where $V(t)$ denotes the volume of an $n$-simplex $t$. The choice $p=-1$ defines the density field $f(v):= f_{-1}(v)$, and we continue the density field onto the rest of the triangulation via a piece-wise linear interpolation. The intuition for the density function $f(v)$ is that, in regions with a high concentration of points (with respect to the standard Lebesgue measure), the Delaunay triangulation tends to build many $n$-simplices with small volume. Vertices in high concentration areas therefore coface several small-volume $n$-simplices, which implies $f(v)$ is larger in high concentration regions than in lower concentration regions. \par 

While performing a superlevel set filtration on $f(v)$ is perfectly valid, in this work we elect to instead perform a sublevel set filtration on the ``inverse density field" \begin{align}
    \ell(v) := f_{1/2}(v) = \left(\sum_{t \in \Delta(v)}V(t)\right)^{1/2}.
\end{align}

Apart from being computationally more efficient, performing the sublevel set filtration on $\ell(v)$ is equivalent to a superlevel set filtration on $f(v)$, up to some monotonic map between the filtration values. In other words, both level set filtrations start at large ``densities" and end at low ``densities".\par 
While the sublevel sets of the linearly-interpolated inverse density $\ell$ are submanifolds, the PH pipeline requires simplicial complexes. Thankfully, the homology of the submanifold is unaltered by instead considering the flag complex of the subgraph of the Delaunay triangulation formed by vertices in the sublevel set \cite{Pranav_Edelsbrunner_van}. Put differently, the topology can only change when vertices are added to the sublevel set filtration, and so the PH pipeline need only consider when vertices enter and exit the filtration, as the edges only appear when both vertices appear. The flag complex is computationally efficient, since it only requires knowledge of the value $\ell(v)$ and the Delaunay triangulation, which is generally sparse.\par 

The Delaunay triangulation in $(\phi,y)$ coordinates is essentially a triangulation in 2D with one periodic boundary condition, which one can visualize as ``unrolling" the infinite cylinder by cutting along the rapidity axis. To properly account for the periodicity of the $\phi$ coordinate, we computationally leverage a trick of duplicating the point cloud twice over (i.e., if $X$ is the point cloud, generate two more point clouds $X(\phi + 2\pi), \, X(\phi-2\pi)$), compute the standard Delaunay triangulation in 2D Euclidean space, and then remove all vertices from $X(\phi+2\pi), \, X(\phi-2\pi)$ that aren't part of the original point cloud $X$ (along with all corresponding simplices). The area of each triangle in the triangulation is then calculated as a function of $\Delta \phi, \Delta y$ (where $\Delta \phi$ is the proper angular separation), and is therefore invariant under a Lorentz boost along the beam axis.\par

While this process is well-established in the literature and properly accounts for our periodic degree of freedom, one unfortunate consequence is the presence of edge effects along the boundary of the point cloud in the rapidity direction. One can easily see this edge effect artifact in Fig. \ref{fig:ph_pipeline}b: the upper boundary exhibits ``sliver" triangles due to the low density along that direction (these slivers are permitted because the circumcircle extends into a region with no points). The consequence of this edge artifact is that the ``sliver" triangles have a low area, and therefore contribute to a larger $f(v)$ for positive $p$. Thus, the boundary points appear to have a larger $f(v)$ than would be implied by the Lebesgue measure. This edge effect substantially affects the PH pipeline by ``turning on" points and simplices at low values of $\ell(v)$ prematurely early. \par 
There are several different ways to combat this issue, each with their own advantages and disadvantages. One approach is to artificially introduce periodic boundary conditions in the $y$ direction as well, such that the high magnitude rapidity points are less likely to produce ``sliver" triangles. However, given the rapidity direction is non-compact, a choice has to be made regarding where to ``cut" along the rapidity direction so as to introduce the periodicity. This choice is somewhat arbitrary and might in and of itself introduce spurious edge effects. Another approach is simply to perform the DTFE with one periodic boundary condition, and then introduce a rapidity cut. The advantage of this approach is computational efficiency, at the cost of a choice of where to introduce the rapidity cut. Based upon our numerical experiments of which high magnitude rapidity points contribute to the edge artifacts, we elected to choose a rapidity cut $|y|<2$ for our PH pipeline. This rapidity cut functionally implies we compute the DTFE for the full point cloud, but exclude the points that lie beyond the rapidity cut for the PH calculation. While we believe this choice to be appropriate and cogent for our work, we designate future work to assessing other methods of mitigating these edge effects.

\section{Different PH protocols}\label{sec:app_diff_PH}
In this appendix, we discuss alternative PH protocols, in the interest of inspiring and informing future work. \par 
The PH pipeline performed in this work utilized a sublevel set density filtration which explicitly depended upon the Delaunay triangulation. The triangulation of the manifold was necessary to specify when two points should be joined together. However, given access to the distances between any two points (either the three-momenta Euclidean distance or the cylindrical distance in $(\phi,y)$), the Vietoris-Rips (VR) filtration could have been performed. In the VR pipeline two points $i,j$ are connected with an edge whenever $d(i,j) < \varepsilon$, where $\varepsilon$ is the filtration parameter and $d(\cdot,\cdot)$ is the distance function. Constructing the flag complex on the resultant graph yields the Vietoris-Rips complex. Note that in this construction the points are assumed to have existed at the beginning of the construction. This has large implications for the Betti number and the cluster entropy, as it implies the zeroth Betti number begins the filtration at its largest value. \par 
One disadvantage of the VR pipeline is the propensity for large chains of points to merge. This effect is well-documented in the context of single-linkage clustering and can be combatted through other agglomerative schemes like Wald or centroid clustering. However, more complicated clustering schemes that avoid these chaining effect are less amenable to higher-dimensional simplicial homology (e.g., clusters, not points are merged in Wald clustering, and so a notion of 1D intra-cluster homology is difficult to define).\par 
The VR pipeline can be modified to have non-uniform open sets around each point. For example, one can build ellipsoids with principal axes that are point-dependent, or radii of balls that scale both with the filtration value and some local functional \cite{kalisnik2020finding}. \par 
Two frequent limitations of PH is computationally large point clouds and the presence of outliers. While large point clouds have less impact on computing 0D homology, higher-dimensional homology becomes computationally intensive. One workaround is to leverage the robustness of PH through the use of a witness complex, wherein a subset of points are used as landmarks used to ``witness" simplicial complexes that reflect the underlying topology \cite{otter2017roadmap,de2004topological}. The witness complex can also be extended through subsampling methods to be robust with respect to outliers \cite{stolz2021outlier}. \par 

\end{document}